\documentclass[]{pasj01}
\draft 
\Received{}
\Accepted{}

\begin{document}

\title{ A Possible Model for the Long-Term Flares of  Sgr A* }
\author{Toru \textsc{Okuda}\altaffilmark{1}}%
\altaffiltext{1}{Hakodate Campus, Hokkaido University of Education, Hachiman-Cho 1-2, 
  Hakodate 040-8567, Japan }
\email{bbnbh669@ybb.ne.jp}

\author{Chandra B.  \textsc{Singh} \altaffilmark{2,5}}
\altaffiltext{2}{ The Raymond and Beverly Sackler School of Physics and Astronomy, Tel Aviv University, Tel Aviv 69978, Israel}

\author{Santabrata \textsc{Das}\altaffilmark{3}}
\altaffiltext{3}{Indian Institute of Technology Guwahati, Guwahati, 781039, India}

\author{ Ramiz \textsc{Aktar}\altaffilmark{3}}

\author{Anuj \textsc{Nandi}\altaffilmark{4}}
\altaffiltext{4}{Space Astronomy Group, SSIF/ISITE Campus, U. R. Rao Satellite Center, 
Outer Ring Road, Marathahalli, Bangalore 560037, India }

\author{Elisabete M. de Gouveia  \textsc{Dal Pino}\altaffilmark{5}}
\altaffiltext{5}{ Department of Astronomy (IAG-USP), University of Sao Paulo, Sao Paulo 05508-090, Brazil}


\maketitle

\begin{abstract}
 We examine  the effects of magnetic field on low angular momentum
 flows with standing shock around black holes in two dimensions.
 The magnetic field brings change in behavior and location of the shock which results
 in regularly or chaotically oscillating phenomena of the flow.
 Adopting fiducial parameters like specific angular momentum, specific energy and 
 magnetic field strength for the flow around Sgr A*, we find that
the shock moves back and forth in the range 60 -- 170$R_{\rm g}$, 
  irregularly recurring  with a time-scale of $\sim$ 5 days
 with an accompanying more rapid small modulation with a period of 25 hrs
 without fading, where $R_{\rm g}$ is the Schwarzschild radius. 
 The time variability associated with two different periods is attributed 
to the oscillating outer strong shock, together with another rapidly oscillating 
 inner weak shock.  As a consequence of the variable shock location,
 the luminosities  vary roughly by more than a factor of $3$. 
 The time-dependent behaviors of the flow are well compatible with  luminous flares
 with a frequency of  $\sim$ one per day and bright flares
 occurring every $\sim$ 5 -- 10 days in the latest observations by Chandra, 
 Swift and XMM-Newton monitoring of Sgr A*. 
\end{abstract}

\section{Introduction}
Our Galactic Center has been extensively studied in the 
framework of accretion processes because it harbors  a supermassive black hole
 candidate, namely Sgr A*, with unique observational features  which are incompatible
 with the standard thin disk model \citep[ hereafter, SS73 model]{key-55}.
One of the remarkable features of Sgr A* is that the observed luminosity 
is five orders of magnitude lower than that predicted by the SS73 model.
Moreover, the spectrum of Sgr A* differs from the multi-temperature black body
spectrum obtained from the SS73 model and the observational features of Sgr A* can not be explained by the 
SS73 model. 
 Then, two types of theoretical models, namely the
spherical Bondi accretion model without any net angular momentum \citep{key-6} and
the advection-dominated accretion flow (ADAF) models with high angular momentum
in the hydrodynamic regime (HD) (Narayan \& Yi 1994, 1995; Stone et al. 1999;
 Igmenshchev \& Abramowicz 1999, 2000; Yuan et al. 2003, 2004)
have been studied (see Narayan \& McClintock 2008, Yuan 2011 and Yuan \& Narayan 2014
  for review). 
Both Bondi model and ADAF model result in highly advected flows with a low
 radiative efficiency compatible with the observations.
However, in contrast to the simple Bondi model,  the advective accretion flow 
models have been generally successful in explaining well the observations \citep{key-13,key-5,key-67,key-29,key-54}.
 In addition to these hydrodynamical studies, since the early works of shear instability in magnetized
 disks \citep{key-2,key-24}, 
 several multidimensional magnetohydrodynamical (MHD) studies have been performed.  They  show that 
 the magnetic fields play important roles in the mass outflows and the flow structure of the accertion disks 
 \citep{key-31,key-32,key-57,key-27,key-38,key-40,key-61,key-62}.
 
 Multi-wavelength studies of Sgr A* have shown two distinct states in Sgr A*: a quiescent state
 and a flaring state (Genzel et al. 2010 and references therein).
 The observations of Sgr A* showed that the durations of the X-ray and IR flares are typically
 of 1 -- 3 hrs and the flare events usually occur a few times per day and that
 the observed emission at radio and IR flares roughly vary by factors of $1/2$ and 
 $1-5$ (Genzel et al. 2003; Ghez et al. 2004; Eckart et al. 2006; Meyer et al. 2006a,b; Trippe et al. 2007;
 Yusef-Zadeh et al. 2009, 2011). 
 While the observed flare emission at X-ray wavelength varies by more than two orders of magnitude
  with respect to the quiescent state \citep{key-50}.

  There are several numerical MHD simulation  works which attempt to address flare phenomena of
  Sgr A* \citep{key-10, key-18,key-19,key-4, key-52}.
 \citet{key-52}, for instance, considered electron thermodynamical effects in general relativistic 
 magnetohydrodynamical (GRMHD) simulations and modeled the emission by thermal electrons, qualitatively
 reproducing some of the observed features.
 In another work, \citet{key-4} showed that non-thermal electrons from highly magnetized regions
 close to the black hole are accelerated due to magnetic reconnection and could be responsible for the rapid variability associated with X-ray flares. 
Recently, a magnetohydrodynamical model for episodic mass ejection from black holes with subsequent  
multi-wavelength flares from Sgr A* has been proposed in analogy with solar coronal mass ejections
to explain many observations of Sgr A*, including their light curves and spectra \citep{key-63,key-30}.
Another scenario considering rotating, radiating  inflow-outflow solutions (RRIOs) \citep{key-40,key-61,key-29,key-62}
 employed a Markov Chain Monte Carlo fitting 
and provided a first globally consistent picture of the Sgr A*  accretion flow, by linking observations to 
 the simulated accretion flows \citep{key-53}.
 It should be noticed that the above works mainly deal with high angular momentum flow and 
 attempt to explain the rapid flares of Sgr A* with a period of 1-3 hrs. 
 However, the latest observations by Chandra, Swift and XMM-Newton monitoring of Sgr A* 
 over fifteen years show that, in addition to the above rapid flares, 
 flares occur at a rate of $\sim$ one per day, while luminous flares occur
 every 5 -- 10 days (Degenaar et al. 2013; Neilsen et al. 2013, 2015; Ponti et al. 2015).

A low angular momentum flow model is an intermediate case between the Bondi model and the ADAF model.
 While 2D hydrodynamical and magnetohydrodynamical simulations of the low angular momentum flows 
 onto black holes  showed that  the magnetorotational instability (MRI) is very robust in the torus even 
 with a weak magnetic field compared with the case of the hydrodynamical flow and that 
 the matter accretes onto the black hole due to the MRI (Proga \& Begelman 2003a, b).
 Also, the standing shock models of the low angular momentum flow have been investigated and applied to Sgr A* \citep {key-8,key-36,key-12}. 
Motivated by their  works, we  examined the low angular momentum flow model for Sgr A*
 using 2D time-dependent hydrodynamic calculations and discussed the implication of their results on the activity of Sgr A* \citep{key-46,key-45,key-47}.
On the other hand, the observational spectra of Sgr A* show a synchrotron emission 
component which is presumably  driven by the magnetic field around Sgr A* \citep{key-50}.
Therefore, a necessary complementary step to these studies is to examine the 
magnetohydrodynamical accretion flow with  low angular momentum.

 In this paper, we examine general effects of the magnetic field on the standing shock in 2D hydrodynamical
 steady flows, using  a  parameter $\beta_{\rm out}$ of the magnetic field strength defined as the ratio of gas pressure to magnetic pressure at the outer boundary.
Then, adopting fiducial parameters of specific angular momentum, specific energy and magnitude of the magnetic field, 
 we apply this scenario to the long-term flares occurring  at a rate of  $\sim$ one per day and also every
 $\sim$ 5 -- 10 days as found in the latest observations of the supermassive black hole candidate Sgr A*.
 
\section{Numerical Methods}
\subsection{Basic Equations}
 We use the public library software PLUTO given by \citet{key-35}.
 PLUTO provides a modular environment capable of simulating hypersonic flows in multi-dimensional coordinates.
 We use here the  magnetohydrodynamics (MHD) module written as a nonlinear system of conservation laws,
 under an adiabatic assumption :

\begin{equation}
 {\partial \rho\over\partial t} + \nabla\cdot\left(\rho\textbf{v}\right) = 0, 
\end{equation}
\begin{equation}
 \frac{\partial (\rho\textbf{v})}{\partial t} + \nabla\cdot\left[
   \rho\textbf{v}\textbf{v}  -     \textbf{B}\textbf{B}\right] 
 + \nabla p_t = -\rho\nabla\Phi,
\end{equation}
\begin{equation}
 \frac{\partial{E}}{\partial t} + \nabla\cdot\left[
  \left(E + p_t\right)\textbf{v} - 
  \left(\textbf{v}\cdot\textbf{B}\right)\textbf{B}\right] = 
    -\rho\textbf{v}\cdot\nabla\Phi,
\end{equation}
\begin{equation}
 \frac{\partial \textbf{B}}{\partial t} + \nabla\cdot\left(\textbf{v}\textbf{B}
  - \textbf{B}\textbf{v}\right) = 0,
\end{equation}
where $\rho$ is the mass density, $\textbf{v}$ is the fluid velocity, $\Phi$ is the gravitational potential, $\textbf{B}$
 is the magnetic field, $p_t=p+\textbf{B}^2/2$ is the total pressure accounting for thermal ($p$), and magnetic ($\textbf{B}^2/2$) contributions. 
The total energy density $E$ is given by
\begin{equation}
 E = \frac{p}{\gamma - 1} + \frac{1}{2}\rho\textbf{v}^2
                                 + \frac{1}{2}\textbf{B}^2\,
\end{equation}
where an ideal equation of state with specific heat ratio $\gamma$ is used.
 We adopt here a pseudo-Newtonian potential \citep{key-48}
 and use  cylindrical coordinates ($R$, $\phi$, $z$).

\subsection{Magnetic Field Configurations}
 To generate the magnetic field,
 we use the vector potential $\textbf{A}$, that is, $\textbf{B} = \nabla\times\textbf{A}$.
 We consider one simple poloidal magnetic field, same as in \citet{key-51-2}, defined by the potential
\begin{equation}
  \textbf{A} = (A_{\rm R}=0,  A_{\rm \phi}= \frac{A_0 z}{rR}, A_{\rm z}=0 ),
\end{equation}
where $r=\sqrt{R^2+z^2}$.

The magnitude of the magnetic field is scaled using the parameter $\beta_{\rm out}
 =8\pi p_{\rm out} /B_{\rm out}^2$ which expresses the ratio of gas pressure to
magnetic pressure at $R_{\rm out}$ on the equator, so that 

\begin{equation}
  A_0=  \rm{sign }(z)  \left(\frac{8\pi p_{\rm out}}{\beta_{\rm out}} \right)^{1/2} R_{\rm out}^2,
\end{equation}
where  $p_{\rm out}$ and $\textbf{B}_{\rm out}$ are  the gas pressure and the strength of the magnetic field
 at the outer boundary $R_{\rm out}$.
 The components ($B_{\rm R}$, $B_{\rm z}$) of the magnetic field are given by
 $B_{\rm R} = -A_0 R/r^3$  and $B_{\rm z} = -A_0 z/r^3$, respectively.
 In this work, we consider relatively weak magnetic field $\beta_{\rm out} = 1000$ -- $10^5$ and consider the computational domain
over the first and fourth quadrants.

\subsection{Initial Conditions}
 Our aim is to examine time-dependent magnetohydrodynamical flow with standing shock.
 We use 2D steady hydrodynamical flow with the standing shock as  initial conditions of the magnetohydrodynamical
 flow. The initial conditions of the 2D hydrodynamical flow are given by approximate 1.5D transonic solutions.
 
\subsubsection{1.5D Transonic solutions }
 We have 1D stationary adiabatic equations of mass, momentum and energy conservations, 
 under the vertical hydrostatic equilibrium assumption. 
 The assumption requires that the relative thickness $h/R$ of the disk 
 is sufficiently small ($h/R \ll$ 1) and results in

  \begin{equation}
    \frac{h}{R} = \left(\frac{R_{\rm gas} T R}{GM}\right)^{1/2} 
    = 0.043\left(\frac{M}{10M_{\odot}}\right)^{-1/2}\left(\frac{T}{10^{10} \; {\rm K}}\right)^{1/2}\left(\frac{R}{3\times 
10^6\; {\rm cm}}\right)^{1/2} ,
  \end{equation}
where $R_{\rm gas}$, $G$, $M$ and $T$ are the gas constant, the gravitational constant, the mass
 of the accreting object and the gas temperature.
 Then, we solve the above mass, momentum and energy conservation equations to find outer and inner critical points 
 and subsequently evaluate radial velocity  $v_{\rm R}$,  sound speed $c_{\rm s}$,  Mach number $M_{\rm a}$, 
 disk thickness $h$ and temperature $T$ at a given radius $R$.
 These 1.5D transonic solutions give the initial conditions of the 2D hydrodynamical flow.

\subsubsection{Standing shock location and vertical hydrostatic equilibrium assumption}
 The standing shock problems and their applications in the astrophysical context have been originally pioneered 
 and developed by \citet{key-20-1} and Chakrabarti and his co-workers \citep{key-7,key-8,key-9,key-14}.
 For a set of specific angular momentum $\lambda$ and specific energy $\epsilon$,  we analytically obtain 
 the global adiabatic transonic accretion solutions with the shock, by solving the hydrodynamical equations in 1.5D 
 that simultaneously satisfy the Rankine-Hugoniot equations at the shock \citep{key-7}.
 The 1.5D transonic solutions are used for assigning primitive variables at a given outer boundary 
 for set up of the 2D hydrodynamical simulation.

We notice that when the outer boundary in the 2D hydrodynamical simulation is chosen not far from the analytically obtained shock location, 
 the numerical solution agrees well with the analytical one in terms of shock position. 
 However, if the outer boundary is located further away, the difference between the numerical and 
 analytical shock locations becomes significant \citep{key-46,key-47}. 
This difference appears not because of the numerical scheme, but mainly attributed due to the assumption of the vertical hydrostatic equilibrium in analytical approach which is not strictly valid and eventually leads to the incorrect transonic solutions. Although other disk geometry, such as constant height model, is sometimes useful to obtain the agreement with the analytical shock location \citep{key-7-1}, it also depends on the flow conditions. 
 However, the above issue shall not apply to cases of pure 1D shock problems since such cases never need 
the disk height.  The analytical shock location obtained from the 1D transonic solutions agrees well with the
 numerical one derived from 1D hydrodynamical simulation \citep{key-35-2}.

Actually, most of the 2D low angular momentum flows are geometrically thick, typically as $h/R \geq$ 0.3
 as shown in later results. Even the advection-dominated flow with a bit larger angular momentum is highly
 advective, hot and geometrically thick as its intrinsic natures, 
 compared with the cold and geometrically thin Keplerian flows with $h/R \ll$ 1 \citep{key-64}.
 In spite of the insufficient agreement between the analytical and 2D numerical shock locations, 
 the 1.5D transonic solutions give theoretically important informations about the existence of the standing shock
 and  characteristic relations between the shock location, the specific angular momentum $\lambda$ and the specific 
 energy $\epsilon$.

\subsubsection{Initial conditions of  magnetohydrodynamical flow}
  The 1.5D transonic solutions give the initial conditions, that is, density $\rho$, radial velocity $v_{\rm R}$,
  sound speed $c_{\rm s}$, Mach number $M_{\rm a}$, pressure $p$ and temperature $T$ within $\mid z \mid \leq h(R)$
  at a given radius $R$ if the mass accretion rate is specified. 
 In the region of  $\mid z \mid$ ＞ $h(R)$, the variables are set appropriately.  
 Then, we perform the simulation until steady state solutions are obtained.
 Finally, we use the steady state hydrodynamical flow as the initial conditions of the magnetohydrodynamical
 simulation.

\subsection{Boundary Conditions}
  In both cases of the hydrodynamical and the magnetohydrodynamical flows, the outer radial boundary 
  at $R=R_{\rm out}$ is divided into two parts. One is the disk boundary through which matter is entering 
 from the outer flow.
 At the disk boundary ( $-h_{\rm out} \leq z \leq h_{\rm out}$ at $R$ =$R_{\rm out}$), we impose continuous
 inflow of matter with constant variables given by the 1.5D solutions.
 The other is the outer boundary region above the accretion disk. Here we impose free floating conditions and allow
 for outflow of matter, whereas any inflow is prohibited.
 At the outer vertical boundary $z=\pm z_{\rm out}$, we also impose the free floating conditions.
 On the rotating axis, all variables are set to be symmetric relative to the axis.
 The inner boundary at $R = R_{\rm in}$ are treated as the absorbing boundary since it is below the last stable
 circular  orbital radius 3$R_{\rm g}$, where $R_{\rm g}$ is the Schwarzschild radius given by $R_{\rm g} = 2GM/c^2$.
 As to the boundary conditions of the magnetohydrodynamical flow, the parameter $\beta_{\rm out}$ of the 
 magnetic field strength is set to be constant at the outer radial boundary.

\section{Numerical Results}

\subsection{Effects of the Magnetic Field on Standing Shock}
 To examine general effects of the magnetic field on 2D hydrodynamical flow with standing shock
 around a black hole with 10 $M_{\odot}$, we use the steady 2D hydrodynamical flow as the initial conditions
 of the magnetized flow.
 A typical set of  parameters of  $\lambda$= 1.65 in units of $2GM/c$ and 
 $\epsilon =6.89 \times 10^{-3}$ in units of $c^2$ is here considered.
 This case has been theoretically examined in detail by \citet{key-7} and
 leads to  inner and  outer critical points at $R$ = 2.80 and 34.6$R_{\rm g}$,
 respectively, and shock location at 18.9$R_{\rm g}$.

 The radial outer boundary $R_{\rm out}$ and the vertical outer boundary $z_{\rm out}$ are taken to be 
 50$R_{\rm g}$ and $R_{\rm in}$ is chosen as $2R_{\rm g}$.
The number of meshes in the cylindrical coordinate ($R, z$) is  $(N_{\rm R}, N_{\rm z})$ = (160, 320). 
 The mesh size is $\Delta R = \Delta z = 0.2R_{\rm g}$ for $0 \leq R \leq 2R_{\rm g}, -2R_{\rm g} \leq 
 z \leq 2R_{\rm g}$, and otherwise $\Delta R = 0.24R_{\rm g}$ and $\Delta z = 0.32R_{\rm g}$.

 Following the steps in subsection 2.3, we simulate this case and get 2D steady state hydrodynamical flow.
  Here, we obtain the shock location $R_{\rm s} = 26.8R_{\rm g}$ on the equator which is larger than 
 the analytical shock location 18.9 $R_{\rm s}$.
 Furthermore, to check the relation between the shock location and the outer boundary used,
 we examined other cases of $R_{\rm out}$ = 30 and 40$R_{\rm g}$ and found the shock locations
 to be 17.3 and 21.1$R_{\rm g}$, respectively. The 1.5D transonic solutions give the relative disk thickness $h/R$ 
 = 0.5 -- 0.6 at 30$R_{\rm g} \leq R \leq 50R_{\rm g}$. Therefore, the condition ($h/R \ll 1$) for 
 the vertical hydrostatic equilibrium is not sufficiently satisfied. 
 However, the case of the smallest boundary radius 30$R_{\rm g}$
 gives the best numerical value of the shock location in agreement with the analytical one.

 Next, using the 2D hydrodynamical steady solutions as the initial conditions,  we solve the time-dependent  2D  
 magnetized flow, by changing  the parameter $\beta_{\rm out}$. 
 Depending on $\beta_{\rm out}$, we represent  states of the flow into three categories: 
 (a)  steady state in weak magnetic fields, (b) quasi-periodic state in intermediate magnetic fields and
 (c) chaotically variable state in strong magnetic fields.
 Fig.~1 shows the time-variations of the shock location $R_{\rm s}$ at the equator for various
 $\beta_{\rm out}$, where the magnitude $\mid \textbf{B}_{\rm out} \mid $ of the magnetic field 
 $\propto  {\beta_{\rm out}}^{-1/2}$.
 In category (a), the magnetic field is very weak as $\beta_{\rm out}$ = $10^5$ and $10^9$,
 but the shock locations are 28.4 and 26.8$R_{\rm g}$ for $\beta_{\rm out}$=  $10^5$ and $10^9$,
 respectively, and increase only a bit with increasing magnetic field, because even a slight increase of the 
 magnetic pressure pushes out the centrifugally supported shock.
  In category (b), the magnetic field is stronger as $\beta_{\rm out}$ = $10^4$ than that in category (a). 
 In category (c), the shock phenomena are very complicated and we find sometimes multiple shocks or
 no shock during the evolution. Although the shock location irregularly oscillates roughly between 
  30 $\leq R_{\rm s} \leq$ 50, the shock seems sometimes to exceed the outer radial boundary. 
 Apart from this outer shock, another inner shock  occurs occasionally and interacts with the outer one.
 Due to these reasons, the shock locations in (c) are denoted by the symbol of filled circle.

\begin{figure}
    \begin{tabular}{ccc}
      
      \begin{minipage}{0.3\linewidth}
        \centering
        \includegraphics[keepaspectratio, scale=0.25]{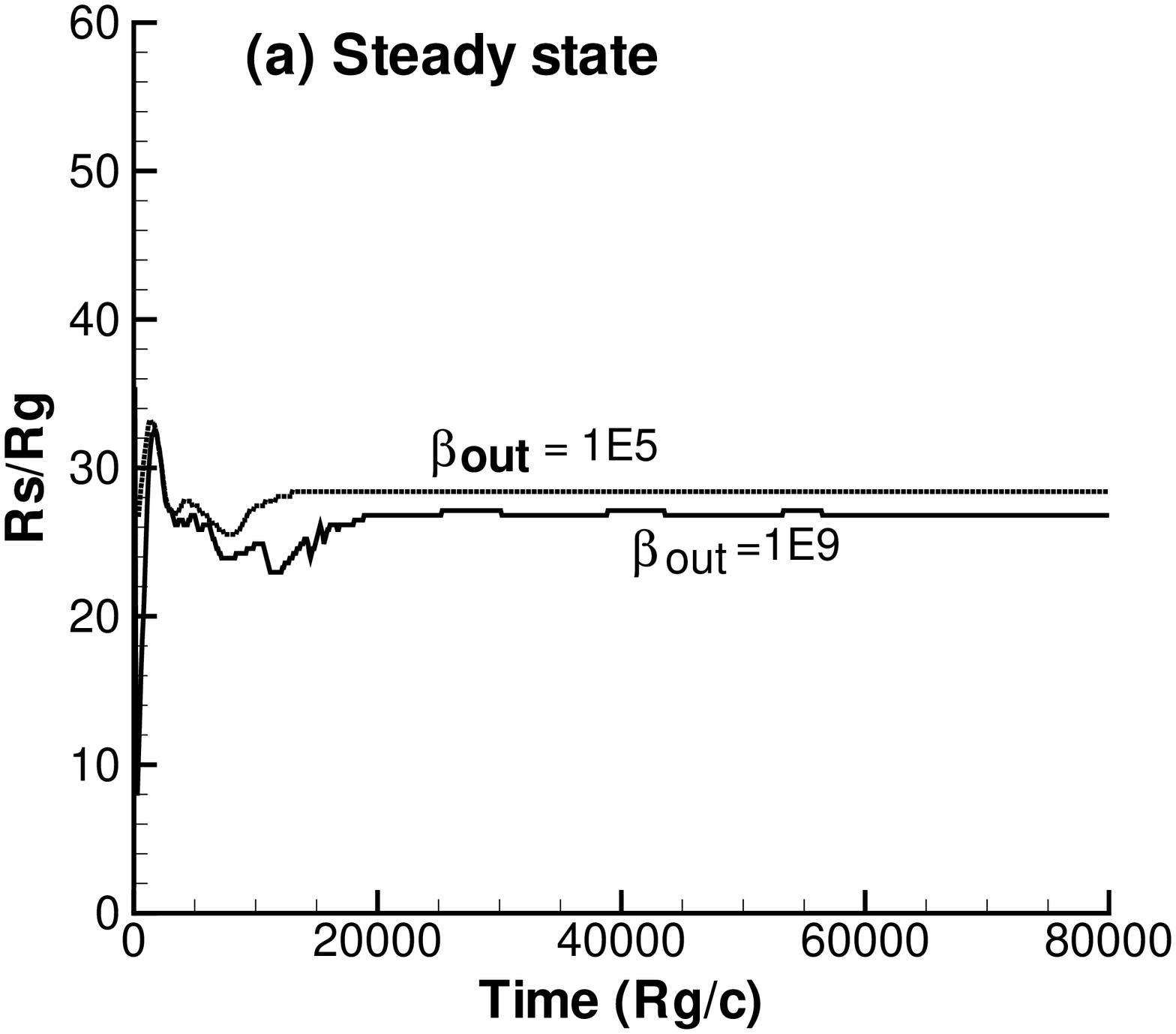}
        \label{ラベル1}
      \end{minipage} 
      
      \begin{minipage}{0.3\linewidth}
        \centering
        \includegraphics[keepaspectratio, scale=0.25]{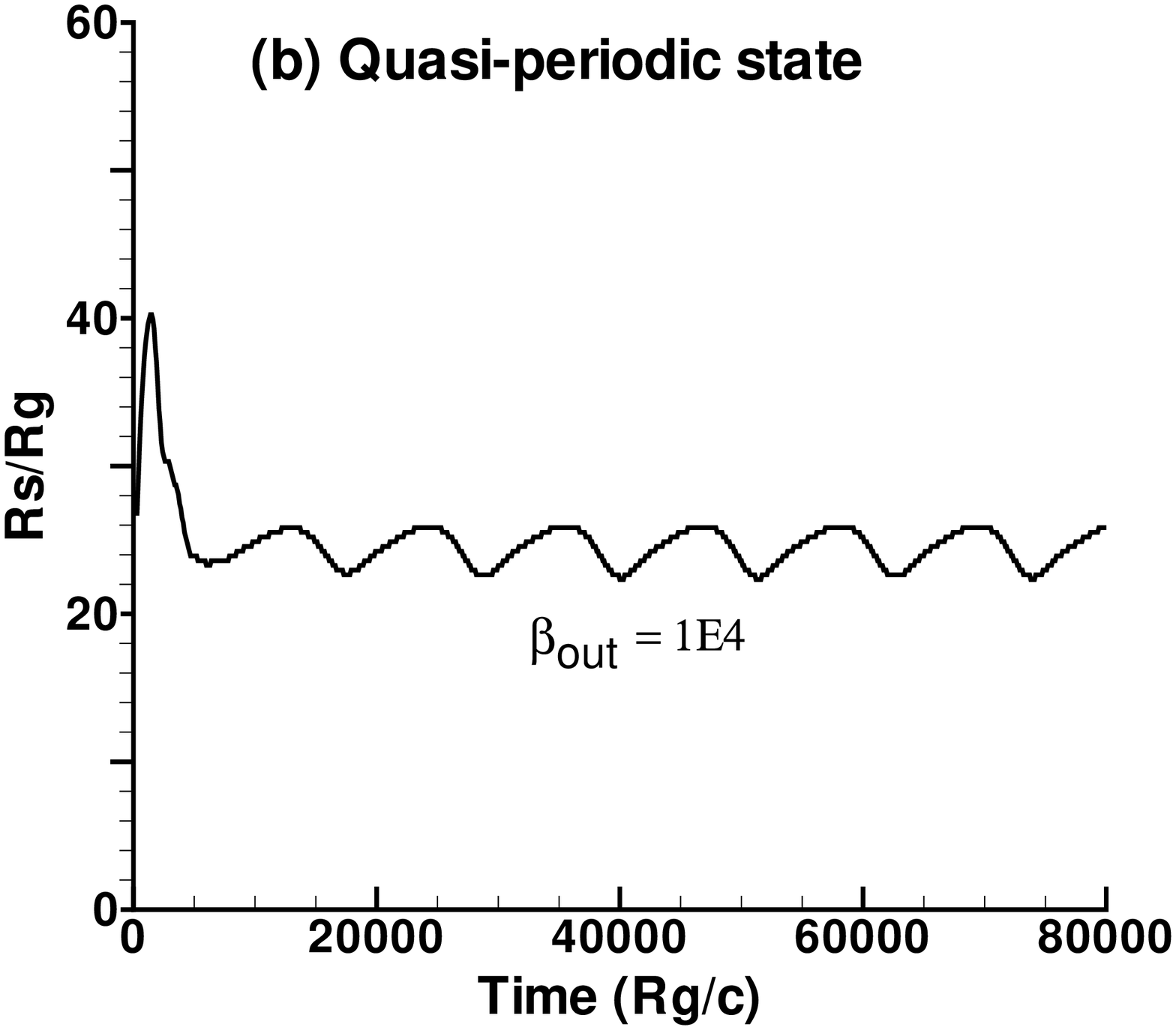}
        \label{ラベル2}
      \end{minipage} 
    
  \begin{minipage}{0.3\linewidth}
        \centering
        \includegraphics[keepaspectratio, scale=0.25]{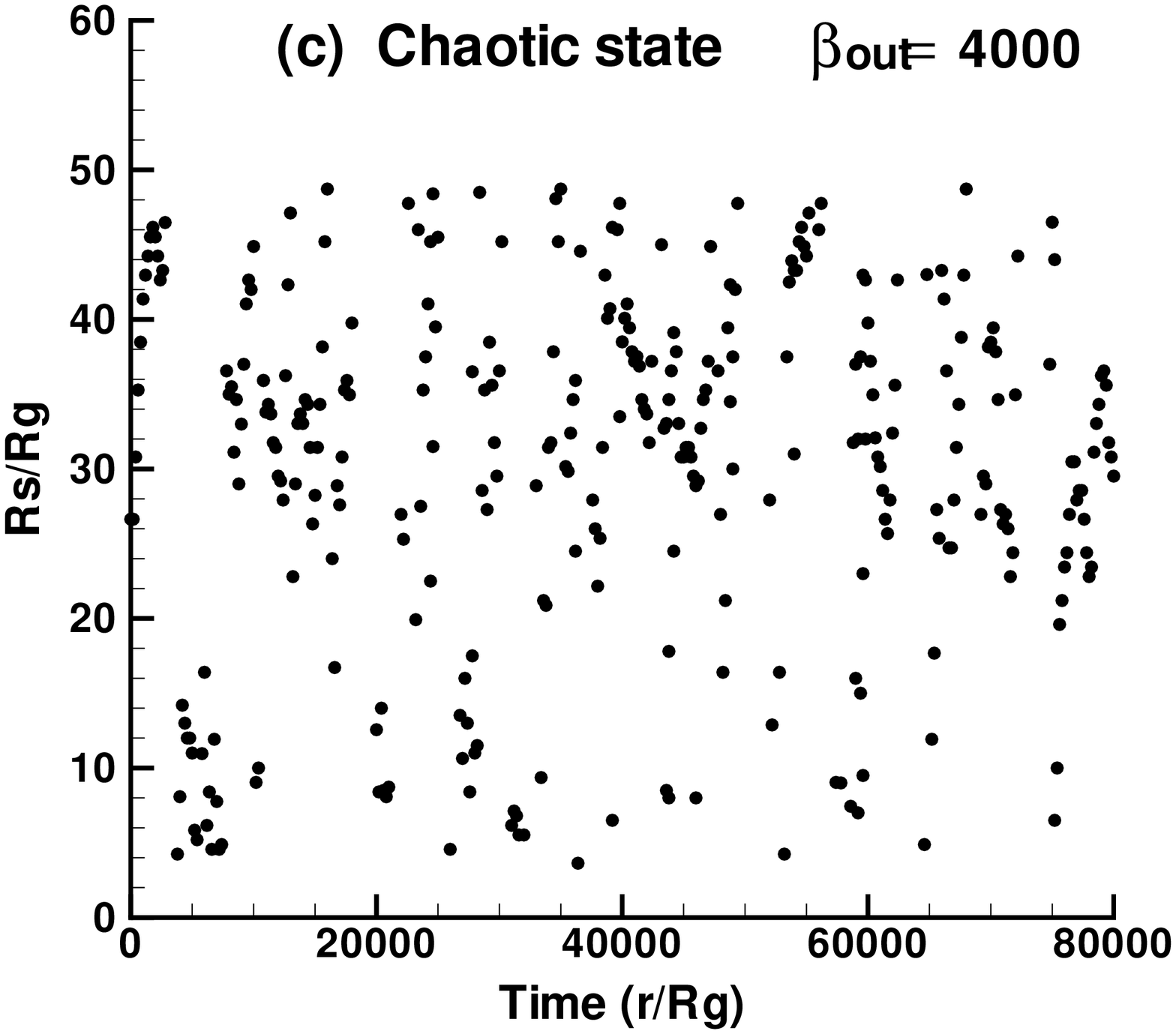}
        \label{ラベル3}
      \end{minipage}
    \end{tabular}
\caption{Time-variations of the shock location $R_{\rm s}$ at the equator for flows with  $\lambda$= 1.65 and
  $\epsilon =6.89 \times 10^{-3}$ for the case of $R_{\rm out}$ = 50$R_{\rm g}$. Depending on the parameter $\beta_{\rm out}$ of the magnetic field strength, the states of the flow are categorized into three types: (a)  steady 
 state in weak magnetic fields, (b) quasi-periodic state in intermediate magnetic fields and (c) chaotically
 variable state in strong magnetic fields. In (a), the shock locations for $\beta_{\rm out}$ = $10^5$ and
  $10^9$ are attained 28.4 and 26.8$R_{\rm g}$, respectively. In (c), the shock locations are shown
 by the filled circle because there exist multiple shocks or no shock during the evolution.
 }
  \end{figure}

 We check whether the flow is subject to the magnetorotational instability (MRI) and 
 whether we are able to resolve the fastest growing MRI mode or not.
 The stringent diagnostics of space resolution for the MRI instability has been examined in 
  3D magnetized flow \citep{key-24-2}.  Therefore, its application to our 2D magnetized flow may be 
 limited to some extent.
 The critical wavelength of the instability mode is given by $\lambda_{\rm c} = 2\pi v_{\rm A}/\sqrt{3}\Omega$,
 where $v_{\rm A}$ and $\Omega$ are the Alfven velocity and the angular velocity \citep{key-3,key-24-2}. 
 A criterion value $Q_{\rm x}$ of the MRI resolution is defined by
 \begin{equation}
    Q_{\rm x}= \frac{\lambda_{\rm c}} {\Delta x},
  \end{equation}
 where $\Delta x$ is the mesh sizes $\Delta R$ and $\Delta z$ in the radial and vertical directions, respectively.
 When $Q_{\rm x}  \gg$ 1, the flow is unstable against the MRI instability, otherwise the flow is stable.
The analyses of  our flows show that $Q_{\rm R} <$ 1 and   $Q_{\rm z} <$ 1 over most region of the flow 
 except for the funnel region along the rotational axis in category (a), and  $40 > Q_{\rm R} > 5$ and 
 $50 > Q_{\rm z} > 10$ near the equatorial plane in categories (b) and (c), indicating that we are able to resolve 
 the MRI in later categories.
 We also calculate the normalized Reynolds stress $\alpha_{\rm gas} = \langle\rho v_{\rm R} \delta v_{\rm \phi}\rangle/\langle 
 p_{\rm g}\rangle$ and the normalized Maxwell stress $\alpha_{\rm mag} = \langle 2B_{\rm R}B_{\rm \phi}\rangle/\langle B^2\rangle$ 
 which are space-averaged over a region near the equator and are time-averaged  over a final duration time
 of the evolution.  Here, $\alpha_{\rm gas}$  and  $\alpha_{\rm mag}$ are roughly  0.06 -- 0.03 and 0.06 -- 0.3,
 respectively, for  $\beta_{\rm out}$ =  4000.

 \begin{figure}
 \begin{center}
     \includegraphics[width=86mm,height=66mm,angle=0]{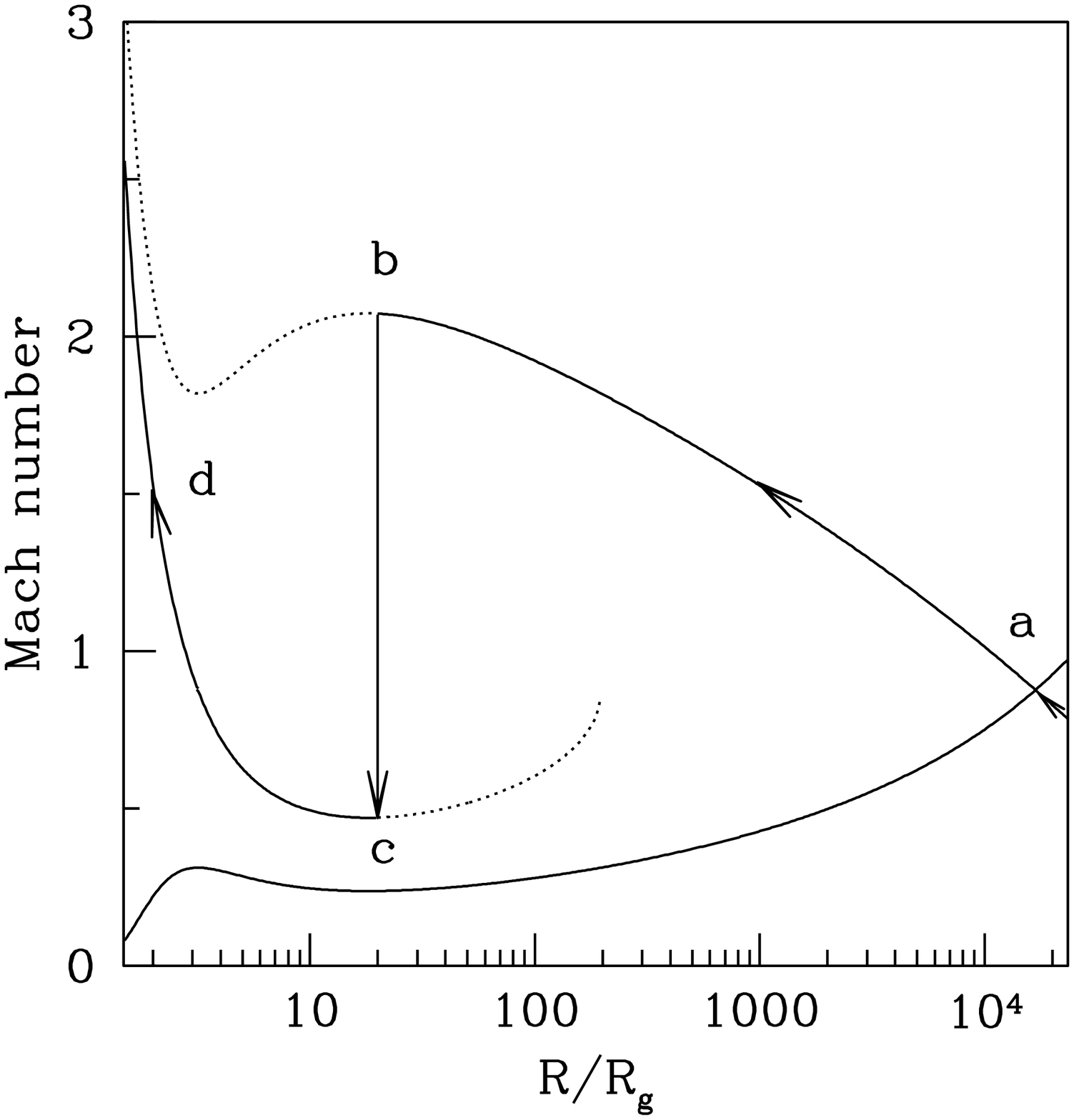}
     \label{fig1}
 \end{center}
 \caption{ Analytical transonic flow  for the parameters of $\lambda$ = 1.35, $\epsilon = 1.98 \times 10^{-6}$
  and $\gamma$ = 1.6 under the assumption of vertical hydrostatic equilibrium of the flow. Vertical arrow
  represents the location of shock transition. Overall arrows indicate the direction of flow motion towards
 the black hole.
}
 \end{figure}

\begin{table*}
\centering
\caption{Parameters of the specific angular momentum $\lambda$, the specific energy $\epsilon$,
the adiabatic index $\gamma$ and the mass accretion rate  $\dot M$, and flow variables of the density
 $\rho_{\rm out}$, the radial velocity $v_{\rm out}$, the temperature $T_{\rm out}$, the relative disk thickness 
 $(h/R)_{\rm out}$, the Keplerian angular momentum $(\lambda_{\rm K})_{\rm out}$ at the outer radial boundary 
 $R_{\rm out}= 200R_{\rm g}$ for Sgr A*  and mesh sizes $\Delta R/R_{\rm g}$, $\Delta z/R_{\rm g}$ 
 used in models A and B.}
\vspace{3mm}
\begin{tabular}{@{}ccccccccccc} \hline
&&&&&&&&& \multicolumn{2}{c}{${\rm Mesh\; sizes}$} \\

$\lambda$ & $\varepsilon$& $\gamma$& $\dot M$
 &$\rho_{\rm out}$ & $v_{\rm out}$ & $T_{\rm out}$ & $ (h/R)_{\rm out}$ 
 & $(\lambda_{\rm K})_{\rm out}$  & $(0 \leq \frac{R}{R_{\rm g}} \leq 2,\; \frac{|z|}{R_{\rm g}} \leq 2)$
 & $({\rm otherwise})$ \\
 ($2GM/c$) & ($c^2$) &  &  ($M_{\odot}\:{\rm yr}^{-1}$) & (${\rm g}\; {\rm cm}^{-3}$) & ($c$) &  ($ \rm K$)
 & &  ($2GM/c$) & ($R_{\rm g}$) & ($R_{\rm g}$) \\
 \hline
  1.35  & 1.98E-6 & 1.6 & 4.0E-6 & 5.87E-19 
 &-0.0498 & 2.55E9 & 0.432 & 10.0   &  0.2  & 0.495 \\
 \hline
\end{tabular}
\end{table*}

 \subsection {Application to the Long-Term Flares of Sgr A*}
 \subsubsection{Setup of the Flow Parameters}
  Here, we consider a supermassive black hole with $M= 4\times 10^6 M_{\odot}$ for Sgr A*.
 Based on the assumption that the Wolf-Rayet star ${\rm IRS}$ 13 $ \rm E3$ is the dominant
source of accreting matter onto Sgr A* and a stellar wind temperature 
 $T_{\rm wind}$ = 0.5 or 1.0 keV, \citet{key-36} estimated a net specific angular momentum 
  $\lambda = 1.68-2.16$,  Bernoulli constant $\epsilon =1.98 \times 10^{-6}$ -- 3.97$ \times 10^{-6}$
 and mass accretion rate $\dot M$ = (2 -- 4) $\times 10^{-6}M_{\odot}$ yr$^{-1}$ for the accretion flow
 around Sgr A*.
 Referring to this work and \citet{key-46}, we consider here a set of parameters  $\lambda$ = 1.35, 
  $\epsilon = 1.98 \times 10^{-6}$ and a mass accretion rate $\dot M$ = 4.0 $\times 10^{-6}M_{\odot}$ yr$^{-1}$
 and examine the time-variations of the magnetized low angular
 momentum flow, focusing on the long-term flares of Sgr A*.

 In Fig.~2, we  show the analytical transonic solution corresponding to the above $\lambda$ and  $\epsilon$, 
where flow after crossing the outer critical point `a' continues to proceed along the supersonic branch `ab' and 
enters the event horizon of the black hole \citep{key-46}, where the outer critical point $R_{\rm a}$
 is 1.68 $\times 10^4 R_{\rm g}$.
However, the flow chooses to jump from point `b' to `c' at  $R_{\rm s} \sim 20 R_{\rm g}$ to become subsonic 
because the entropy generated through the shock is higher compared to that of the supersonic branch.
Subsequently, the flow passes through the inner critical point and becomes supersonic again along `cd' 
before crossing the event horizon.

Since we focus on the long-term variability of Sgr A*,  it is desirable for us to set the standing shock at large
radius.  Taking account of the relation between the numerical shock location and the
 adopted outer boundary radius in subsections 2.3 and 3.1, we set the outer radial boundary at
  $R_{\rm out} = 200 R_{\rm g}$ and determine the primitive variables  $\rho$, $\textbf{v}$ and $p$ at the boundary from the transonic solutions.
  As to the parameter $\beta_{\rm out}$ of the magnitude of the magnetic field, 
  considering of the effects of magnetic field on the standing shock in subsection 3.1, we adopt two cases of 
  $\beta_{\rm out}$= 1000 (model A) and 5000 (model B).

Table 1 shows the model parameters of the specific angular momentum $\lambda$, the specific energy $\epsilon$,
the adiabatic index $\gamma$ and the mass accretion rate  $\dot M$, and the flow variables of the density
 $\rho_{\rm out}$, the radial velocity $v_{\rm out}$, the temperature $T_{\rm out}$, the relative disk thickness 
 $(h/R)_{\rm out}$, the Keplerian angular momentum $(\lambda_{\rm K})_{\rm out}$ at the outer radial boundary $R_{\rm out}= 200R_{\rm g} $ for Sgr A* and mesh sizes $\Delta R/R_{\rm g}, \Delta z/R_{\rm g}$ used in models A and B.
 In our adiabatic flow model, the specific angular momentum $\lambda$ is kept constant everywhere.
 The Keplerian angular momentum $\lambda_{\rm K} \propto R^{1/2}$ and 
 is larger than the constant $\lambda$ in most of regions considered here.

 The computational domain consists of $0 \leq R \leq R_{\rm out} = 200 R_{\rm g}$ and $- z_{\rm out} 
\leq z \leq z_{\rm out}$ with $z_{\rm out} = 200R_{\rm g}$.
The number of meshes is  $(N_{\rm R}, N_{\rm z})$ = (410, 820). The mesh size is $\Delta R = \Delta z =
 0.2R_{\rm g}$ for $0 \leq R \leq 2R_{\rm g}, -2R_{\rm g} \leq z \leq 2R_{\rm g}$, 
 and otherwise $\Delta R = \Delta z = 0.495R_{\rm g}$.

 After the steps described in section 2, we obtain the steady state hydrodynamical flow.
 Fig.~3 shows the profiles of temperature $T$, Mach number of the radial velocity $v_{\rm R}$ and
 the relative disk thickness $h/R$ at the equator (Left)
 and  the 2D density contours log $\rho$ in g cm$^{-3}$ (Right) for the steady hydrodynamical flow. 
The disk thickness is geometrically thick as $h/R \simeq 0.4$.
 The shock is clearly distinguished as a sharp discontinuity at $R= 64.8 R_{\rm g}$.
 In the 2D contours of the density, the shock extends from the equatorial plane 
 obliquely toward the upper stream.
 
  The numerically obtained shock location $R_{\rm s} = 64.8 R_{\rm g}$ at the equatorial plane differs considerably 
 from the analytical value $R_{\rm s} \sim$ 20$R_{\rm g}$ as discussed in subsection 2.3.
 Then, using the 2D steady hydrodynamical flow  as the initial conditions of the magnetized flow,
 we examine the time-variations of the shock location $R_{\rm s}$ and the total luminosity $L$
 of the magnetized flow.

\begin{figure}
 \begin{center}
  \begin{tabular}{c}
  \begin{minipage} {0.5\linewidth}
   \begin{center}
    \includegraphics[width=70mm,height=60mm,angle=0]{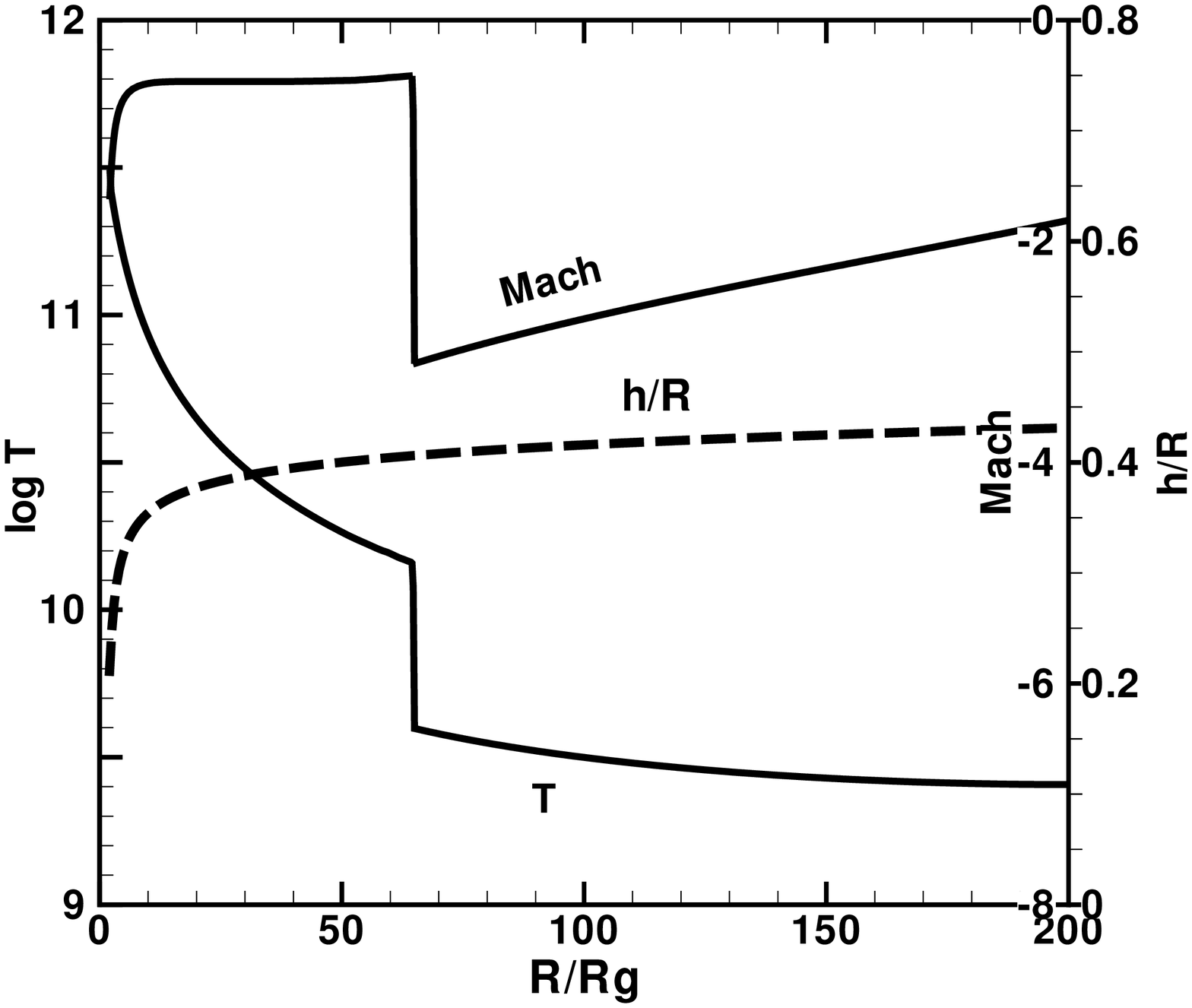}
     \label{fig3-1}
 \end{center}
 \end{minipage}

 \begin{minipage}{0.5\linewidth}
  \begin{center}
   \includegraphics[width=70mm,height=60mm,angle=0]{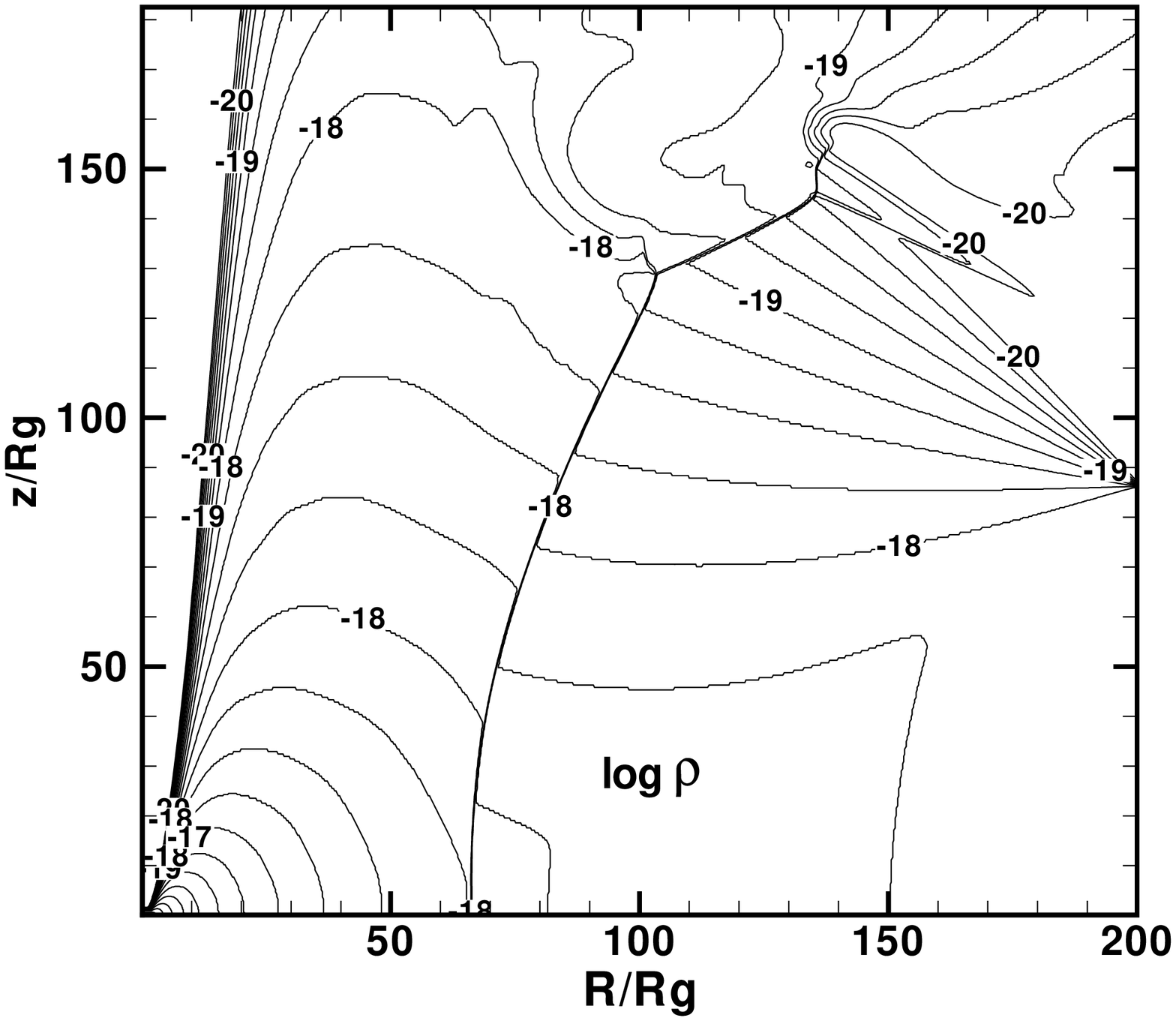}
     \label{fig3-2}
  \end{center}
  \end{minipage}
  \end{tabular}
 \vspace*{5mm}
 \caption{ Profiles of temperature $T$, Mach number of the radial velocity $v_{\rm R}$ and relative disk
 thickness $h/R$ at the equator (Left)
 and  2D density contours log $\rho$ in g cm$^{-3}$ (Right) 
 for the hydrodynamical steady state flow. 
 The shock is clearly distinguished as a sharp discontinuity at $R= 64.8 R_{\rm g}$.
 In the contours, only the first quadrant is presented and 
 the shock extends from the equatorial plane obliquely toward the upper stream.}
 \end{center}
 \end{figure}
 
 The  luminosity $L$ is given by
 \begin{equation}
 L=\int q_{\rm ff} {\rm d} V,
\end{equation}
 where $q_{\rm ff}$ is the free-free bremsstrahlung emission rate per unit volume and here only ion-electron 
bremsstrahlung is considered under a single temperature model and $L$ is integrated over all
 computational zones.  
 If we consider a two-temperature model and a stronger magnetic field as  $\beta_{\rm out}$ = 1 -- 10, 
 the synchrotron radiation  dominates the free-free emission \citep{key-45}
 and equation (10) may be invalid in such synchrotron radiation dominated region.
 However, the adiabatic assumption of the energy equation is considered to be reasonable everywhere
  because even the synchrotron cooling rate is far smaller than the transfer rate of the advected thermal energy.
The mass-outflow rate $\dot M_{\rm out}$ is defined by the total rate of outflow
 through the outer boundaries ($z= \pm ~ z_{\rm out}$) in the z direction,

 \begin{equation}
  \dot M_{\rm out} = \int_{0}^{R_{\rm out}} 2 \pi \rho (R, z_{\rm out}) v_{\rm z}(R, z_{\rm out}) R dR
        - \int_{0}^{R_{\rm out}}2 \pi \rho (R, -z_{\rm out}) v_{\rm z}(R, -z_{\rm out}) R dR ,
 \end{equation}
 where $v_{\rm z} (R,z)$ is the vertical velocity as a function of the coordinate ($R$, $z$).
The outflow rate $\dot M_{\rm R_{\rm out}}$ through the outer boundary ($R=R_{\rm out}$) in the R direction is 
 not included in the above equation. 

 \subsubsection{Evolution of the Magnetized Flow}
 In our models, the centrifugal pressure-supported shock possesses  high temperature and high density
 post-shock matter and is termed as post-shock corona (hereafter PSC) \citep{key-1}. 
The existence of the PSC is a good tracer of the flow evolution in our models.
As is found in subsection 3.1,  we expect  the flow is subject to the MRI also in models A and B.
 Actual  analyses of the models  show that $100 > Q_{\rm R} >$ 10 and  $100 > Q_{\rm z} >$ 20 at
 $R > 6R_{\rm g}$ near the equatorial plane, indicating that we are able to resolve the MRI.
 Hereafter, focusing on model A, we explain the whole evolution of the magnetized flow, because the pattern
 of  time-variability of $L$ and $ R_{\rm s}$ in model B is basically similar to that in model A.
 
 Fig.~4 shows the profiles of the density $\rho$, the angular velocity $\Omega$, the gas pressure $p$, 
 the magnetic pressure $p_{\rm m}$, the normalized Reynolds stress $\alpha_{\rm gas}$ and the normalized Maxwell
 stress $\alpha_{\rm mag}$ for model A, where $\Omega$ and pressure $p$ are given 
 in units of $c^3/GM$ and dyn/cm$^2$, respectively, and the variables are space-averaged between
 $-2R_{\rm g} \leq z \leq 2R_{\rm g} $ and are time-averaged  over the last duration time $1.1 \times 10^7 - 1.2 \times 10^7$ s.
 In the plots of density and angular velocity, their initial values and the Keplerian angular velocity are also shown.
 In spite of the MRI activity during the evolution, 
 the averaged angular velocity is rather distributed along its initial value and is far smaller than the Keplerian one.
 The strong jump at the shock in the initial density is smoothened out  in the averaged density due to
 the irregularly oscillating shock.
   The Maxwell stress is much stronger than the Reynolds stress in the inner region and
  $\mid \alpha_{\rm mag}\mid$ is  $\sim$ 0.1 in the region of $60 R_{\rm g} \leq R \leq 170 R_{\rm g}$, 
 where the shock oscillates. The averaged magnetic pressure is far smaller than the gas pressure
  in the outer region due to the given large $\beta_{\rm out}$ at $R = R_{\rm out}$ but increases 
  toward the inner region.
  These distributions of $\Omega$, $p$, $p_{\rm m}$, $\alpha_{\rm gas}$ and $\alpha_{\rm mag}$ are similar to 
  those at the same radial region in the low angular momentum magnetized flows by \citet{key-51-2}.

\begin{figure}
    \begin{tabular}{cc}
           \begin{minipage}{0.4\linewidth}
        \centering
        \includegraphics[keepaspectratio, scale=0.3]{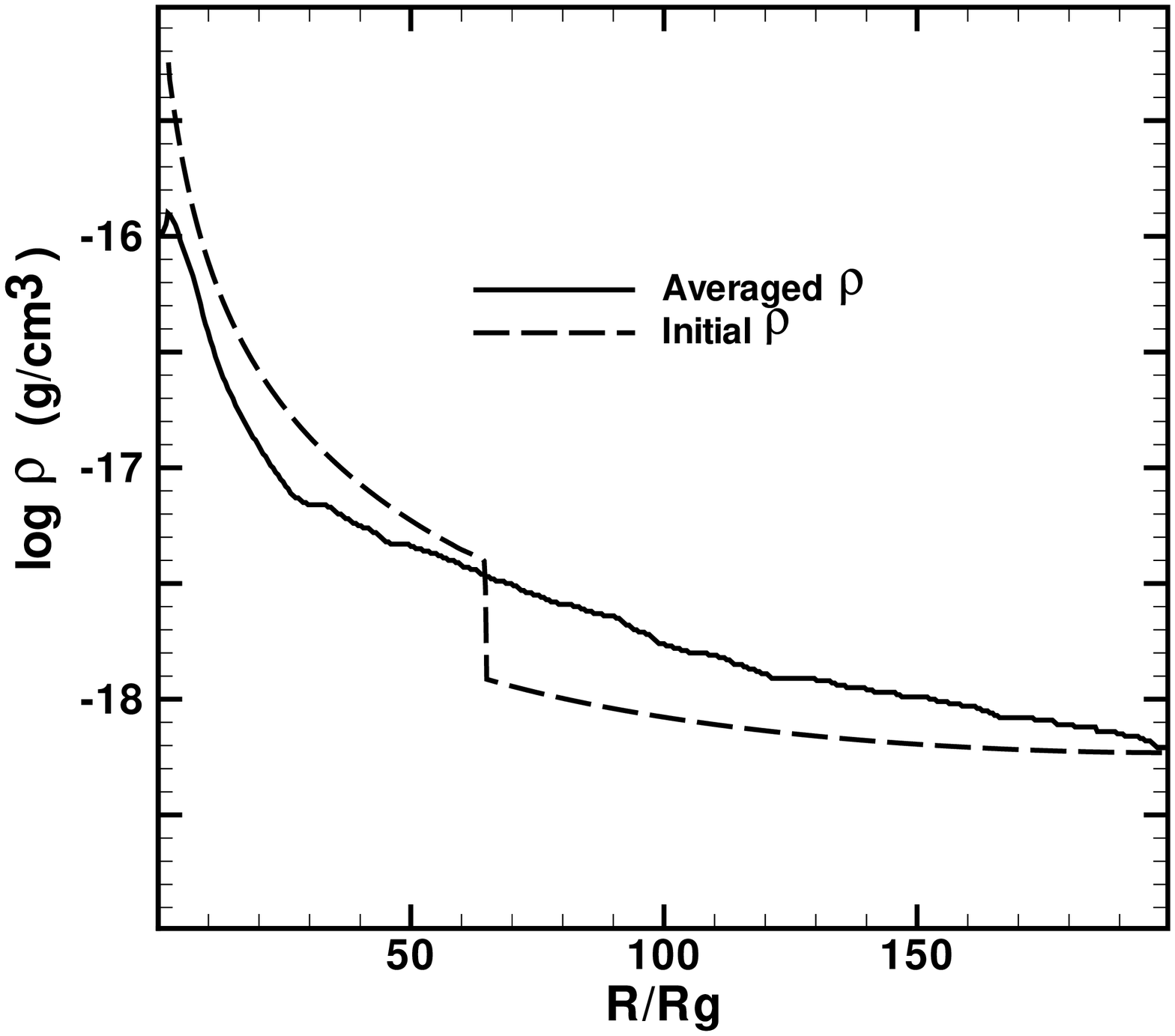}
        \label{ラベル4-1}
      \end{minipage} 
      
      \begin{minipage}{0.4\linewidth}
        \centering
        \includegraphics[keepaspectratio, scale=0.3]{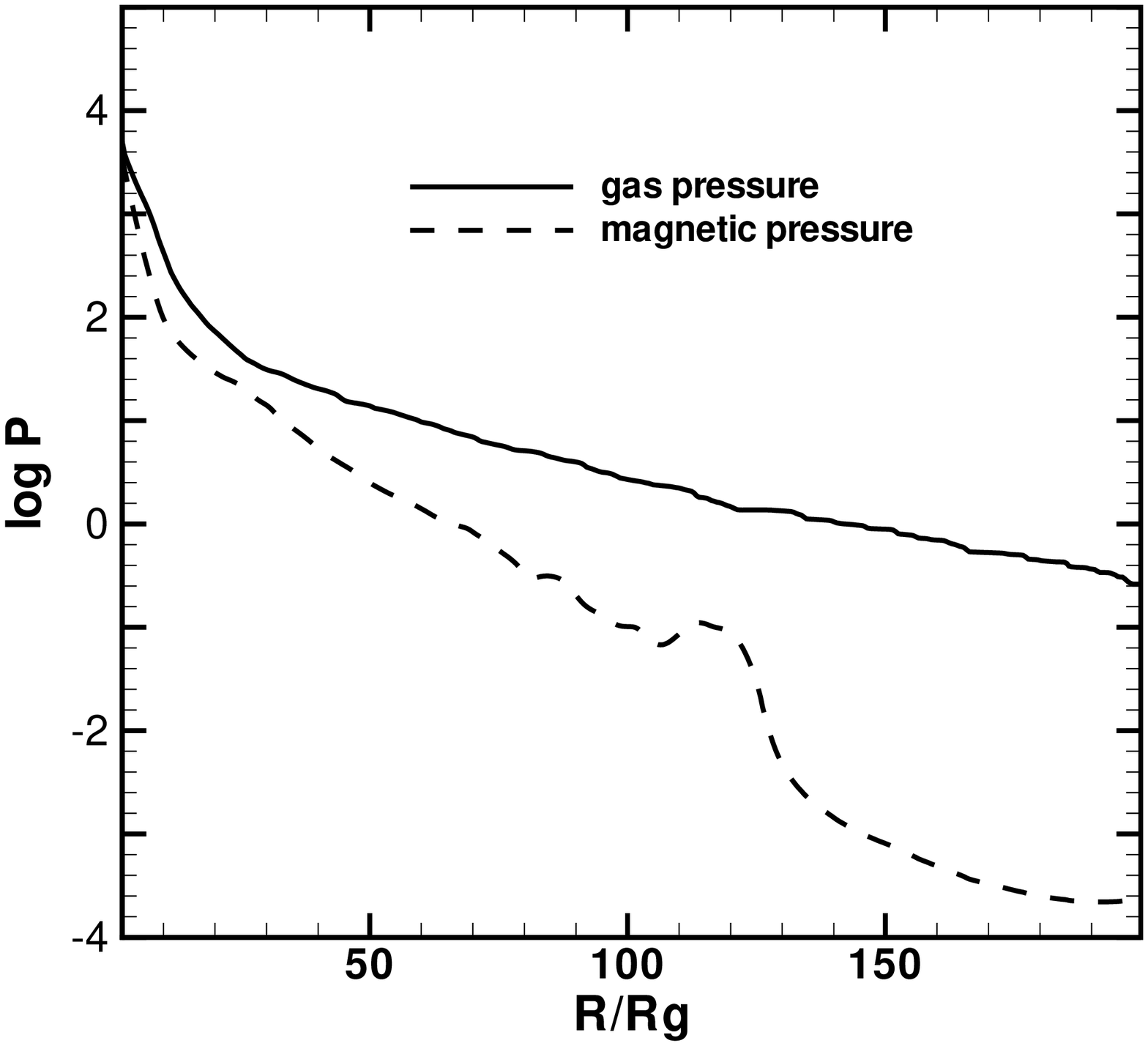}
        \label{ラベル4-2}
      \end{minipage} \\
    
  \begin{minipage}{0.4\linewidth}
        \centering
        \includegraphics[keepaspectratio, scale=0.3]{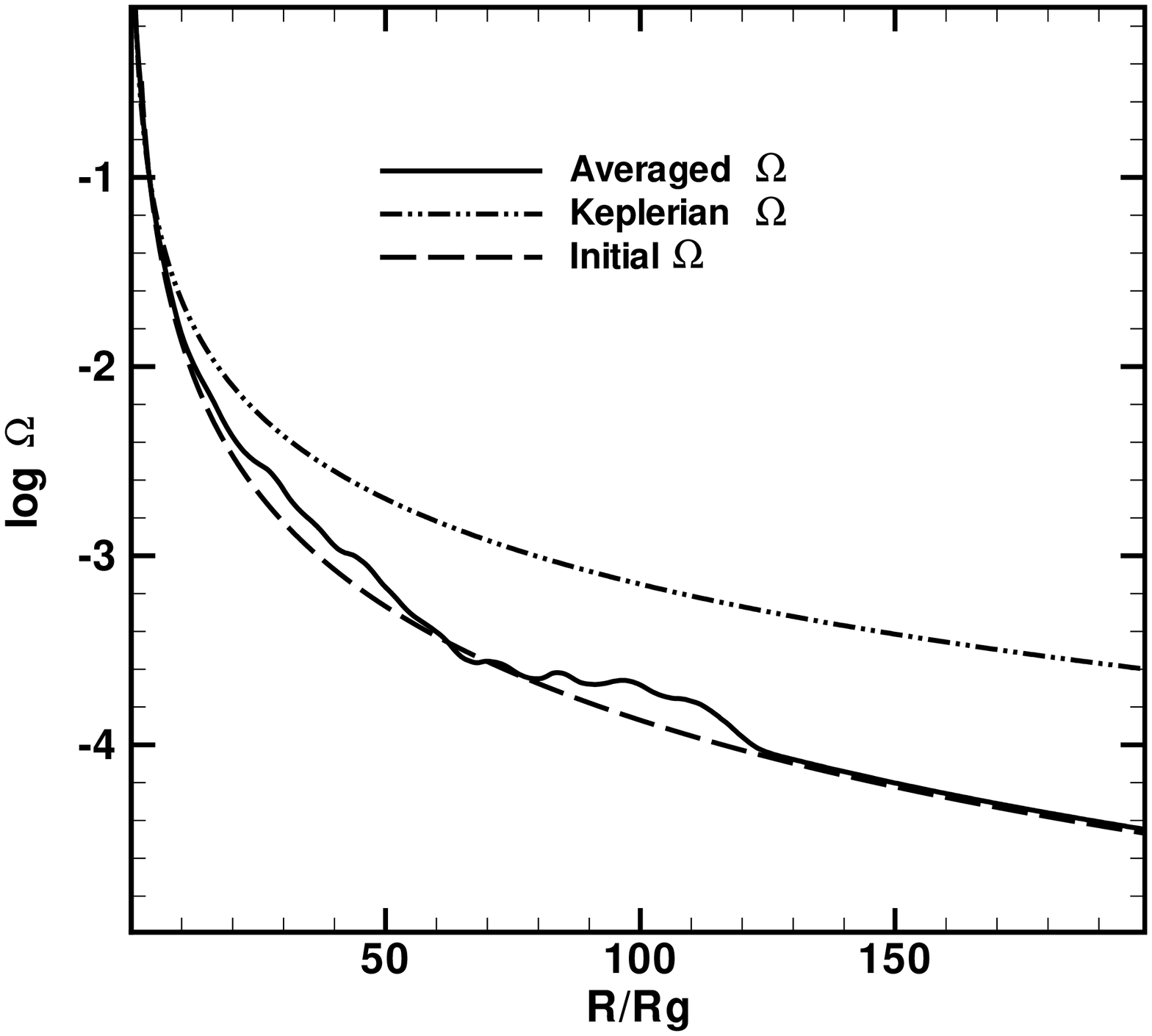}
        \label{ラベル4-3}
      \end{minipage}

  \begin{minipage}{0.4\linewidth}
        \centering
        \includegraphics[keepaspectratio, scale=0.3]{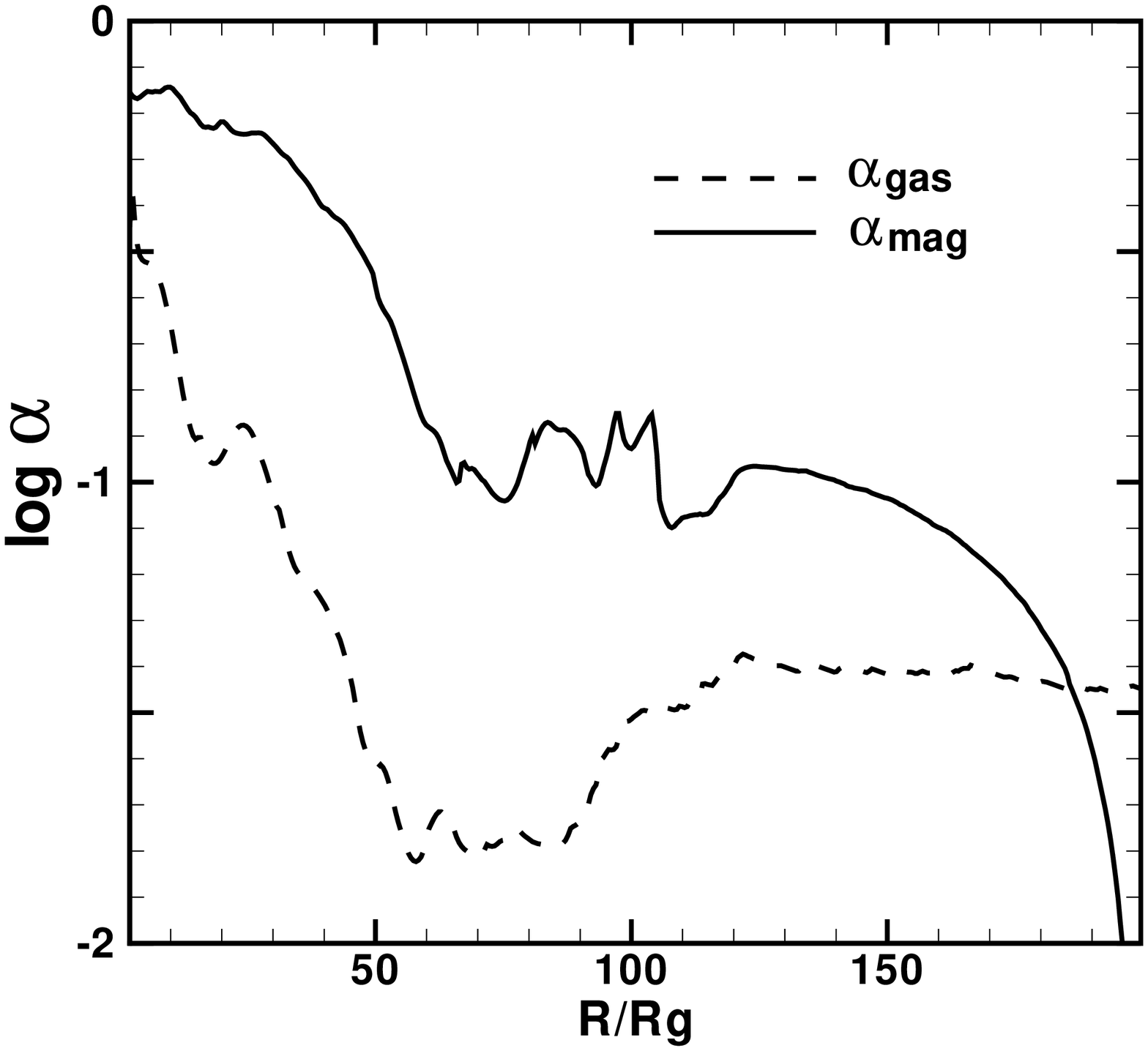}
        \label{ラベル4-4}
      \end{minipage}
    \end{tabular}
          \caption{Radial profiles of the density $\rho$, the angular velocity $\Omega$, the gas pressure $p$, 
           the magnetic pressure $p_{\rm m}$, the normalized Reynolds stress $\alpha_{\rm gas}$ and 
           the normalized Maxwell stress $\alpha_{\rm mag}$ for model A.  The variables are space-averaged
           between $-2R_{\rm g} \leq z \leq 2R_{\rm g} $ and are time-averaged  over the last duration time $1.1 \times 
           10^7 - 1.2 \times 10^7$ s. In the plots of density and angular velocity, the initial values and
           the Keplerian angular velocity are also shown. 
           }
  \end{figure}

After a transient initial phase, the magnetic field is amplified rapidly by the MRI
and  the MHD turbulence  near the equatorial plane, as well as by the advection of the magnetic field lines to
the inner boundary. The flow and shock structures  are
 considerably different from the initial hydrodynamical flow  and show 
very asymmetric features above and below the equatorial plane due to the magnetic field. 
The luminosity and the shock location vary  by more than  a factor of  3 in models A and B.
Fig.~5 shows the time sequence of the velocity vectors and 2D contours of the density $\rho$, 
the magnetic field strength $\mid \textbf{B}^2 \mid$ and the ratio $\beta$ of gas to magnetic
 pressure at times $ t= 2\times 10^5$ (a), $4\times 10^5$ (b), $8\times 10^5$ (c), $1.2\times 10^6$ (d)
 and $1.6\times 10^6$ (e) s (left to right) in model A, 
where the velocity vectors and the magnetic field strength $\mid \textbf{B} \mid$ are shown in unit vector and in 
unit of Gauss (G), respectively. 
 To compare  with the  observations of Sgr A*, henceforth we use time unit of seconds.
 The luminosity and the shock location at (a), (b), (c), (d) and (e) are  $2.9\times 10^{34},
 3.6\times 10^{34}, 3.1\times 10^{34}, 1.6\times 10^{34}$ and $1.3\times 10^{35}$ erg s$^{-1}$ and
 108, 122, 140, 187 and 128$R_{\rm g}$, where the luminosities at (d) and (e) are minimal
  and maximal, respectively, in the entire time-evolution of the luminosity (see Fig.~7).
 The final time  $1.6\times 10^6$ s is indicated as (e) in the curves of $L$ and $R_{\rm s}$ in Fig.~7.
During the times depicted, the MRI grows and stabilizes.
 As a result, MHD turbulence develops near the equator.

 A high magnetic blob is formed in the inner region within the PSC region.
 Here, we designate the high magnetic blob as a spherical bubble-like
 shape with high magnetic field strength which is clearly found in the third and fourth 
 panels of Fig.~5. The high magnetic blob is distinguished from the broader PSC region behind 
 the centrifugal pressure-supported shock.
 The PSC region behind the shock  has high density and high temperature but the magnetic field just
 behind the shock is not so strong as that in the high magnetic blob.
 The blob goes forward with increasing magnetic field strength, continues to expand diffusively and 
 obliquely across the equator up to $R \sim 140R_{\rm g}$ at time (d) 
 and then fades out as a filament-like feature at time (e).
This morphological evolution reflects directly on the time evolution 
of the luminosity and the shock location in Fig.~7. 

 Focusing on (d) and (e) of Fig.~5 where the luminosity becomes minimal and maximal, respectively,  
 we examined the magnitude $\mid \textbf{B} \mid$ of the magnetic field. Fig.~6 shows the contours of 
 $\mid \textbf{B} \mid$ in the inner region for model A.
 We find here that 50 G $\geq \mid \textbf{B} \mid \geq$ 20 G in (d) and
 30 G $\geq \mid \textbf{B} \mid \geq$ 3 G in (e)  at 20$R_{\rm g} \geq R \geq 5R_{\rm g}$ on the equator,
 while 10 G $\geq \mid \textbf{B} \mid \geq$ 0.1 G in (d) and
 1 G $\geq \mid \textbf{B} \mid \geq$ 0.1 G in (e)  at 200$R_{\rm g} \geq R \geq 50R_{\rm g}$, respectively,
 where $\beta_{\rm out}$=1000 corresponds to the magnetic field strength $\mid \textbf{B} \mid \sim$ 0.1 G 
 at the outer radial boundary.
 In each contour of Fig.~6,  there exist two distorted central masses of high magnetic field
 $\mid \textbf{B} \mid \geq $ 60 G. They are very unstable and yield filamentous projections of the magnetic field
 which develop into a spherical bubble-like shape (high magnetic blob) in the outer region.

 The behavior of  $\beta$ in the bottom panel of Fig.~5 is 
 compatible with the evolution of the magnetic field shown in the same figure.
 These panels of  Fig.~5 show how the shock wave evolves due to the effect of the magnetic pressure. 
 During the initial phases (a) -- (b),  due to the MRI activity, $\beta$ becomes low as $\sim 10$
 in the high magnetic blob region which extends to $R$ = 50 and 70$R_{\rm g}$, respectively,  
 in (a) and (b) but $\beta \sim$ 100 -- 1000 outside the high magnetic blob. 
 
 On the other hand, $\beta$ is very small in the funnel region along the rotational axis, making these regions nearly magnetically dominated. Accordingly, the gas is strongly accelerated along the rotational axis, compared with the initial 
 hydrodynamical model.
 The outflow region is roughly separated into two regions: one of  a high velocity jet in the funnel region 
 along the rotational  axis and another one with a wind at the vertical outer boundaries. 
 The jet  has velocities between $0.1 - 0.8 c$ within the funnel region 
 and the wind has roughly the escape velocity $v_{\rm e} \sim 0.07c$ outside the funnel region.
 The averaged mass-outflow rate from the jet and the wind occupies 40 -- 50$\%$ of the input accretion rate 
 in models A and B and the remainder gas falls into the black hole. 
 The mass outflow rate of the jet is comparable to that of the wind.
 Since the magnetic field is not axisymmetric to the equator, the flow features are different above and below 
the equator, but the qualitative behavior is similar in both regions.
 After $t \sim 2\times 10^6$ s, the MRI activity settles to a nearly  stable state, but the 
luminosity and the shock location are strongly variable with irregular oscillation and  large amplitude (see Fig. 7). 
 
\begin{figure}
    \begin{tabular}{ccccc}
      
      \begin{minipage}{0.2\linewidth}
        \centering
        \includegraphics[keepaspectratio, scale=0.25]{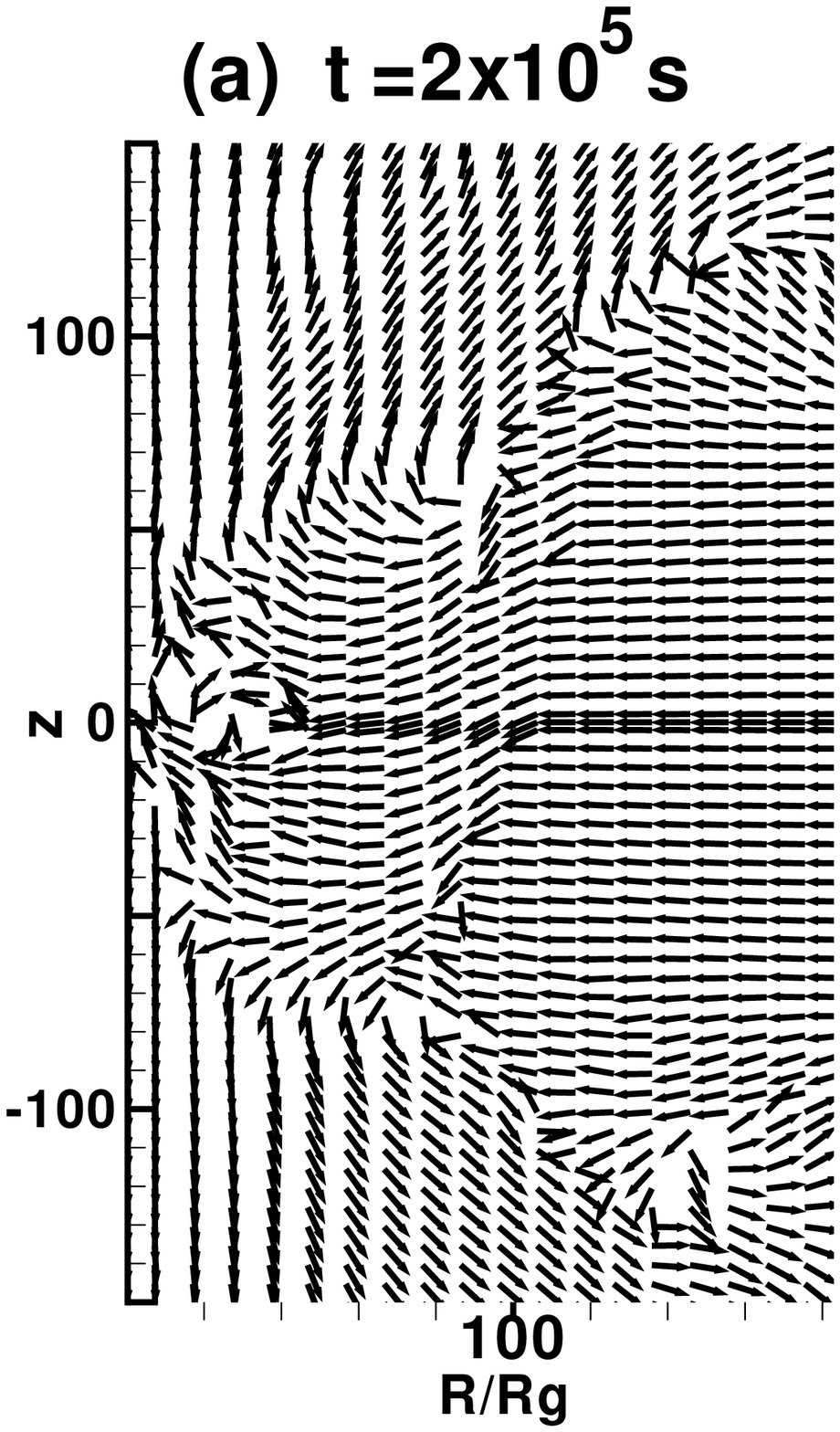}
        \label{ラベル1}
      \end{minipage} 
      
      \begin{minipage}{0.2\linewidth}
        \centering
        \includegraphics[keepaspectratio, scale=0.25]{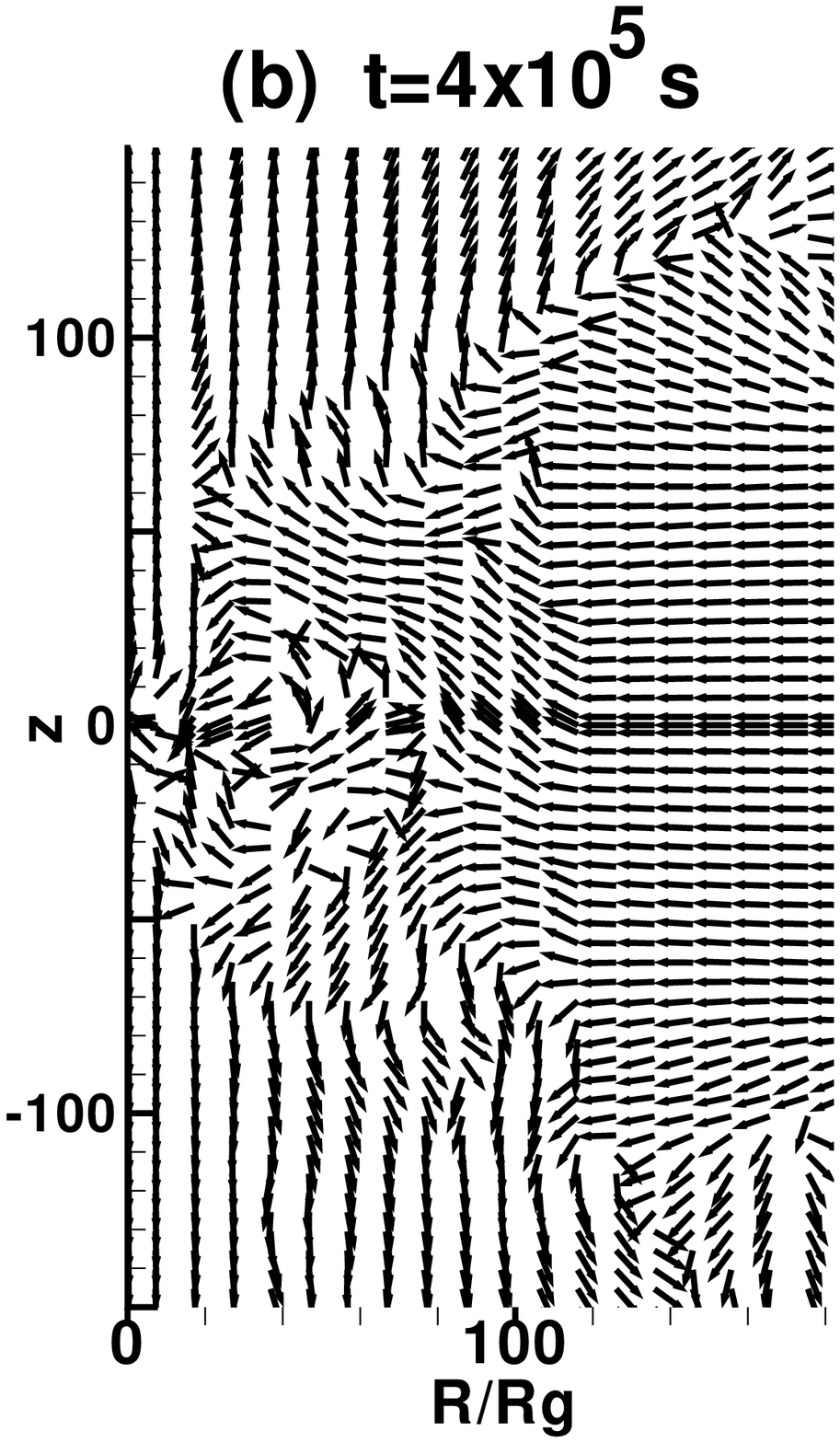}
        \label{ラベル2}
      \end{minipage} 
    
  \begin{minipage}{0.2\linewidth}
        \centering
        \includegraphics[keepaspectratio, scale=0.25]{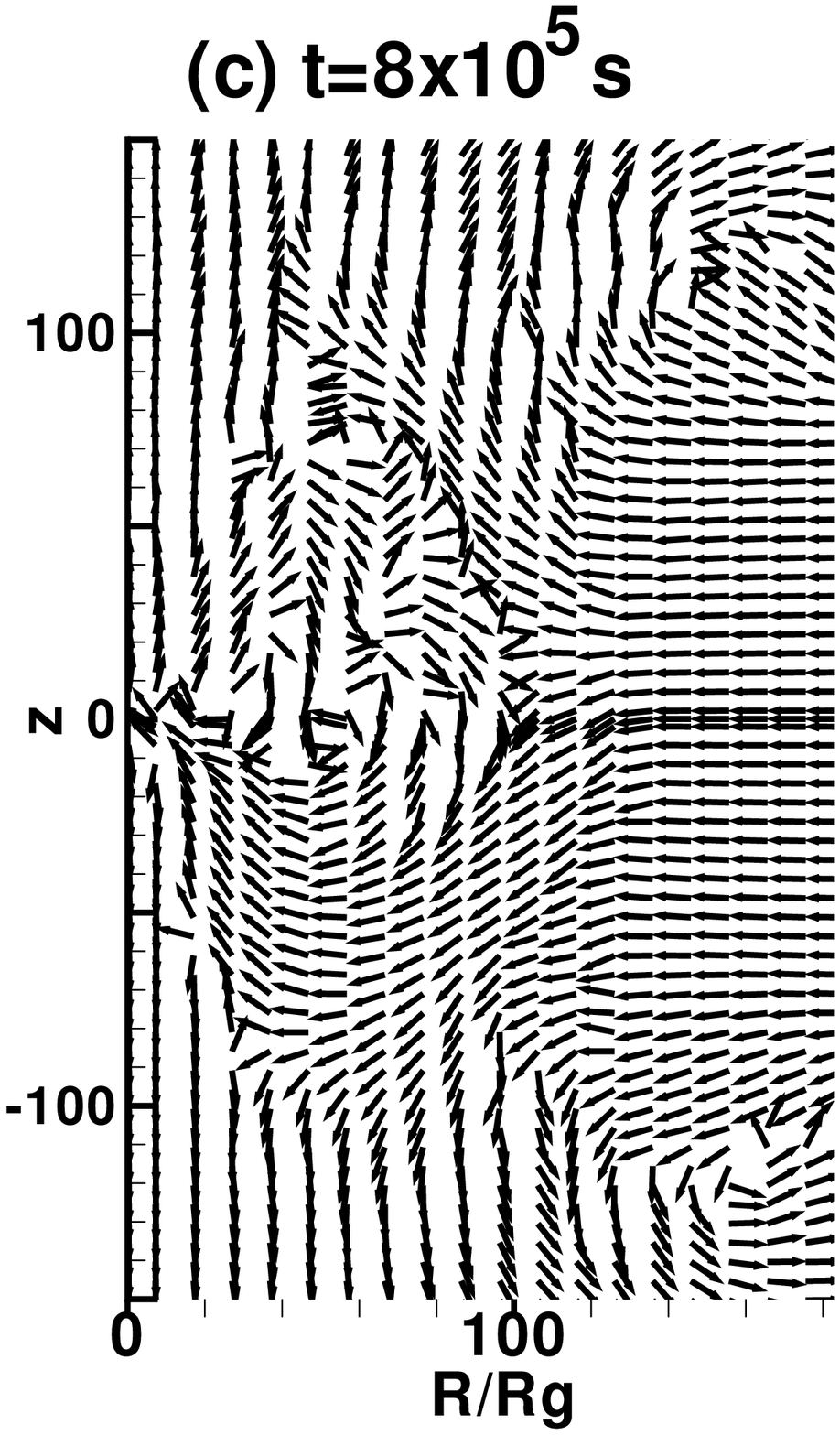}
        \label{ラベル3}
      \end{minipage}

  \begin{minipage}{0.2\linewidth}
        \centering
        \includegraphics[keepaspectratio, scale=0.25]{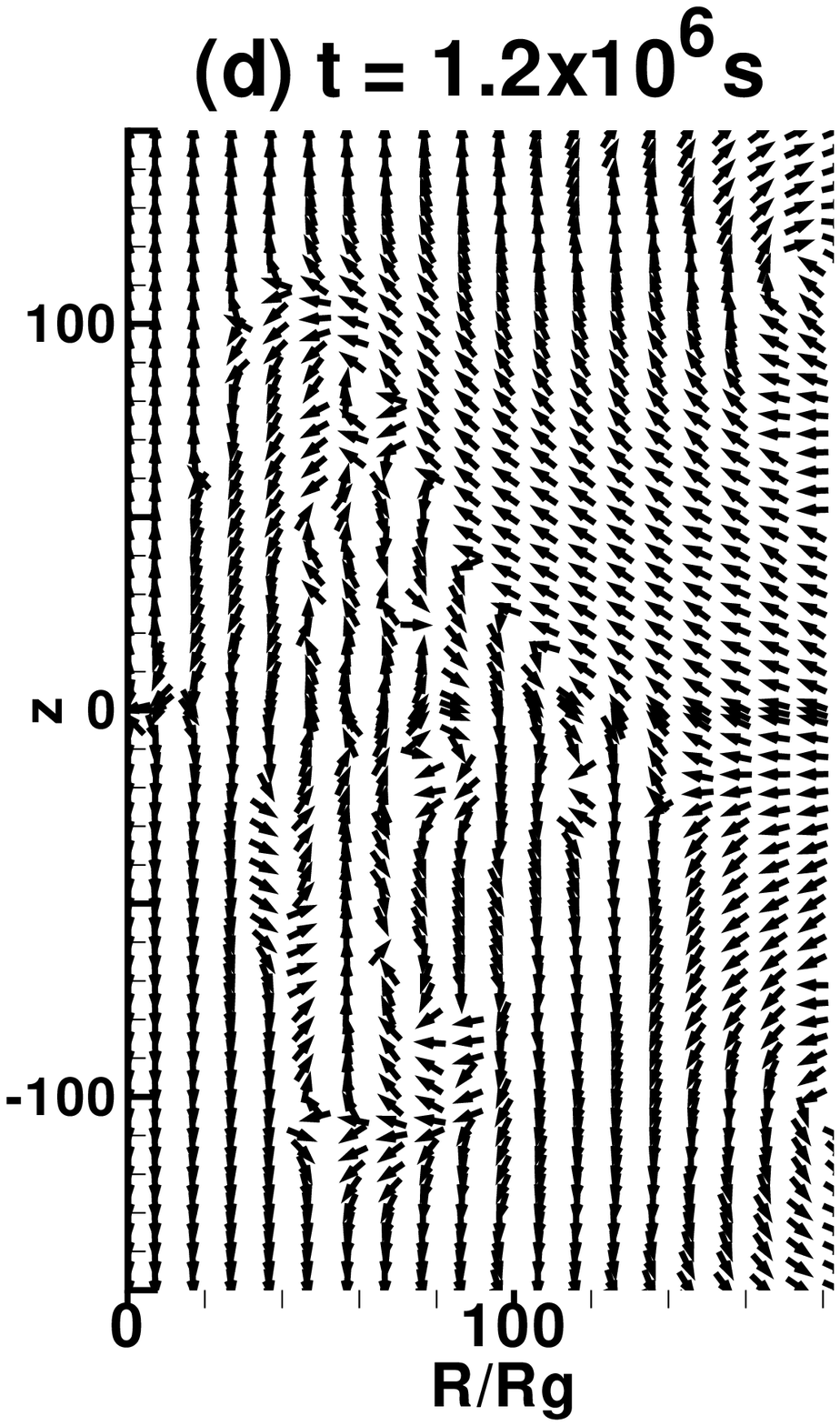}
        \label{ラベル4}
      \end{minipage}

 \begin{minipage}{0.2\linewidth}
        \centering
        \includegraphics[keepaspectratio, scale=0.25]{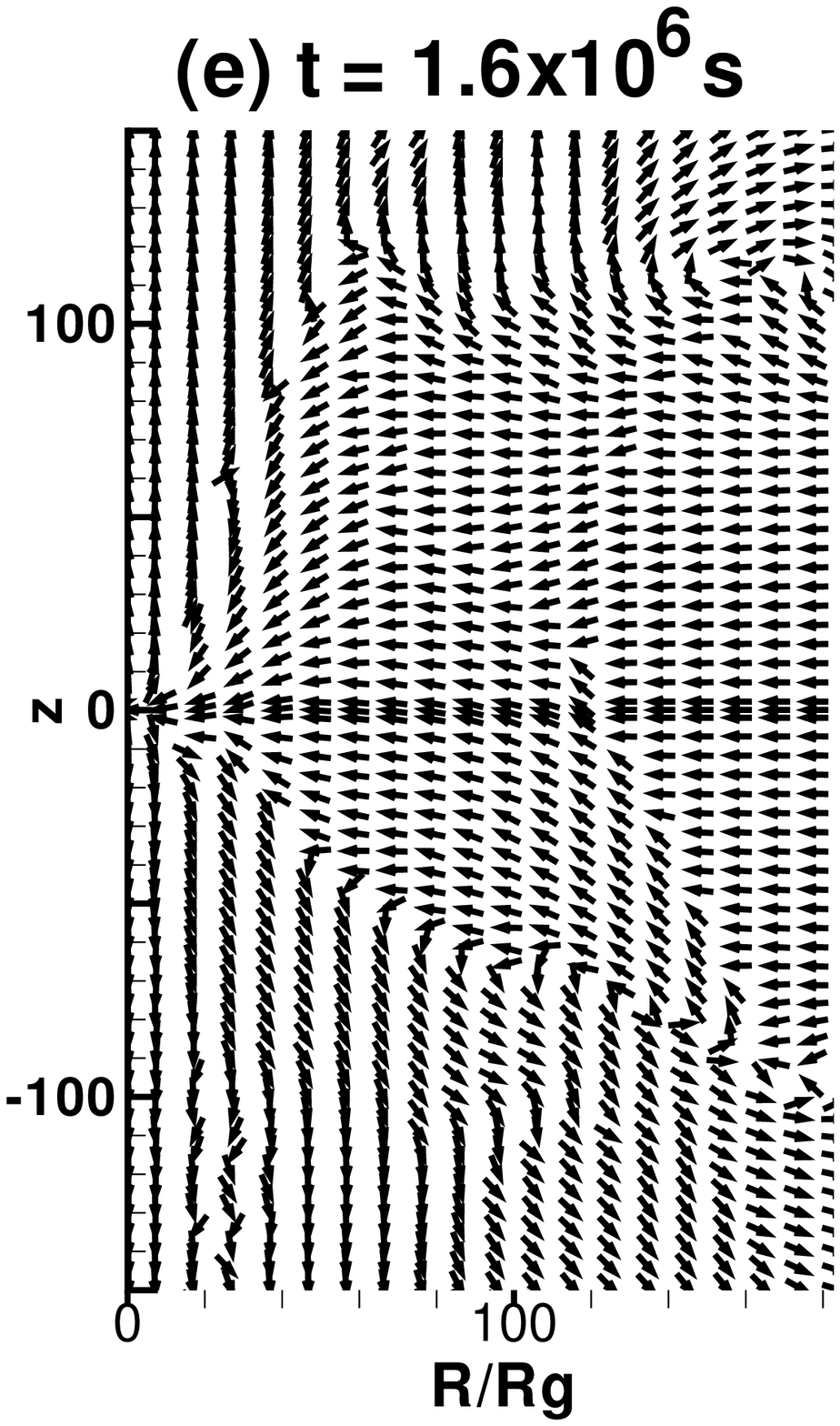}
        \label{ラベル5}
      \end{minipage}\\

      \begin{minipage}{0.2\linewidth}
        \centering
        \includegraphics[keepaspectratio, scale=0.25,angle=-90]{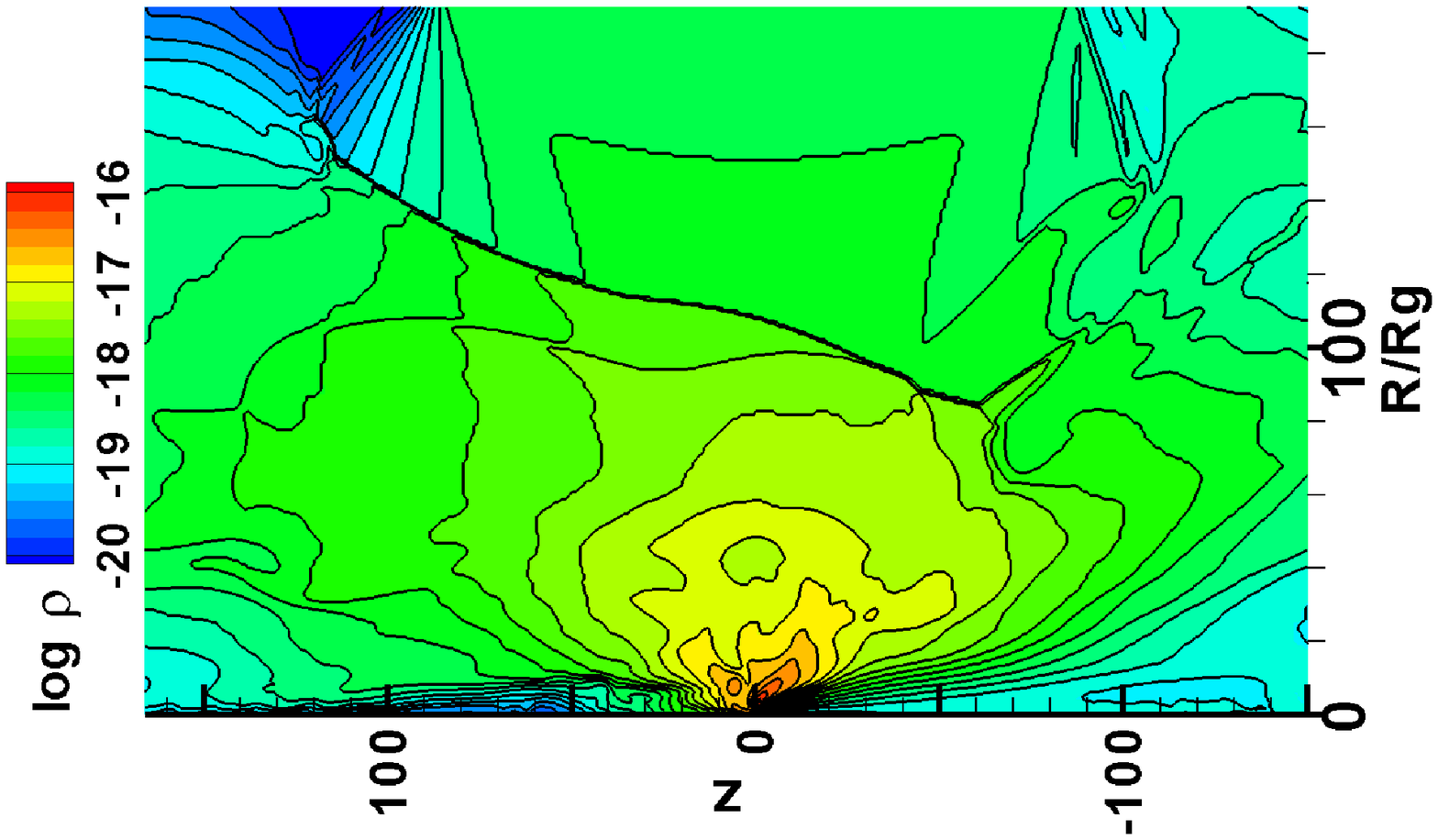}
        \label{ラベル6}
      \end{minipage} 
     
      \begin{minipage}{0.2\linewidth}
        \centering
        \includegraphics[keepaspectratio, scale=0.25,angle=-90]{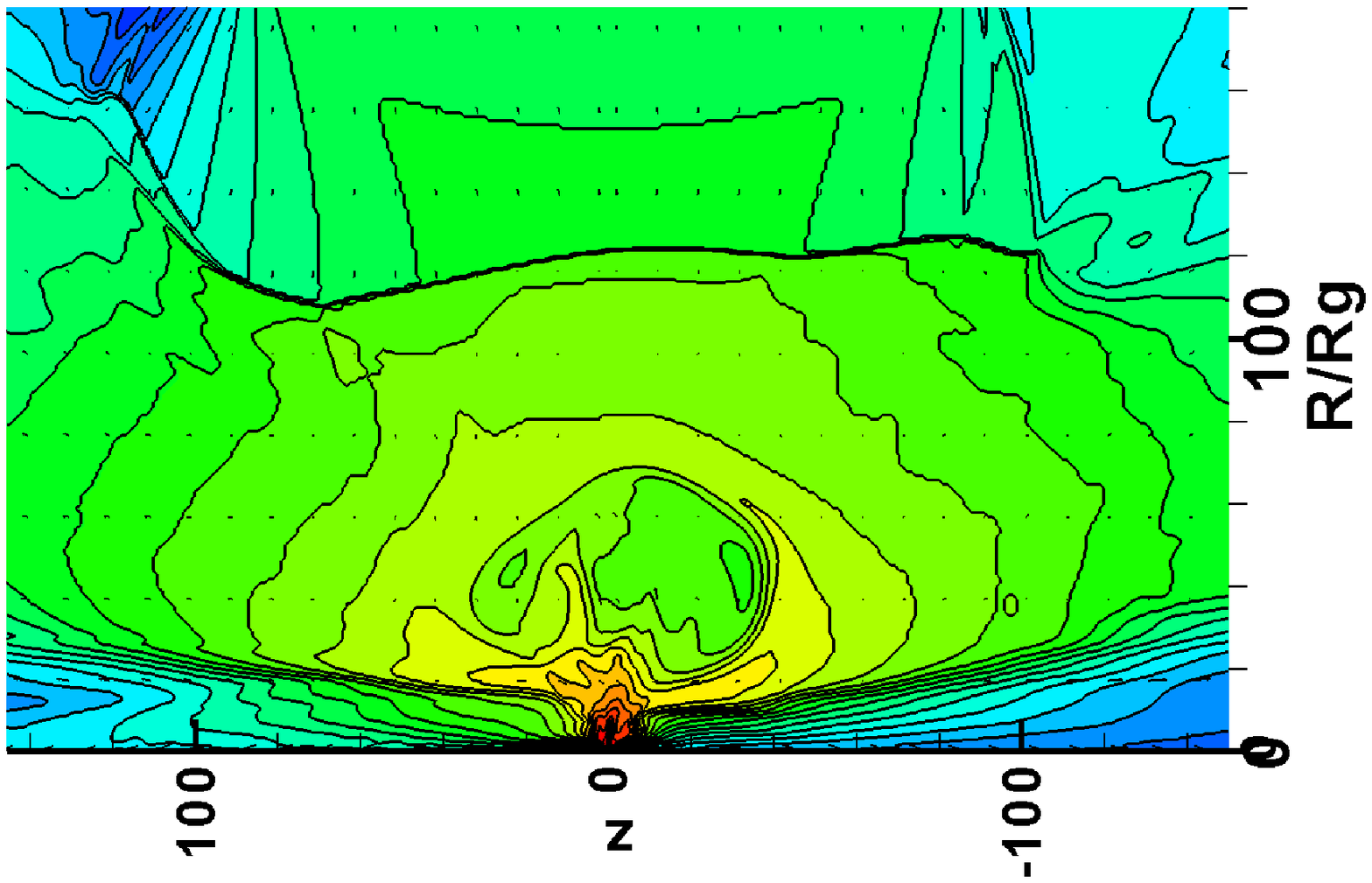}
        \label{ラベル7}
      \end{minipage} 

  \begin{minipage}{0.2\linewidth}
        \centering
        \includegraphics[keepaspectratio, scale=0.25,angle=-90]{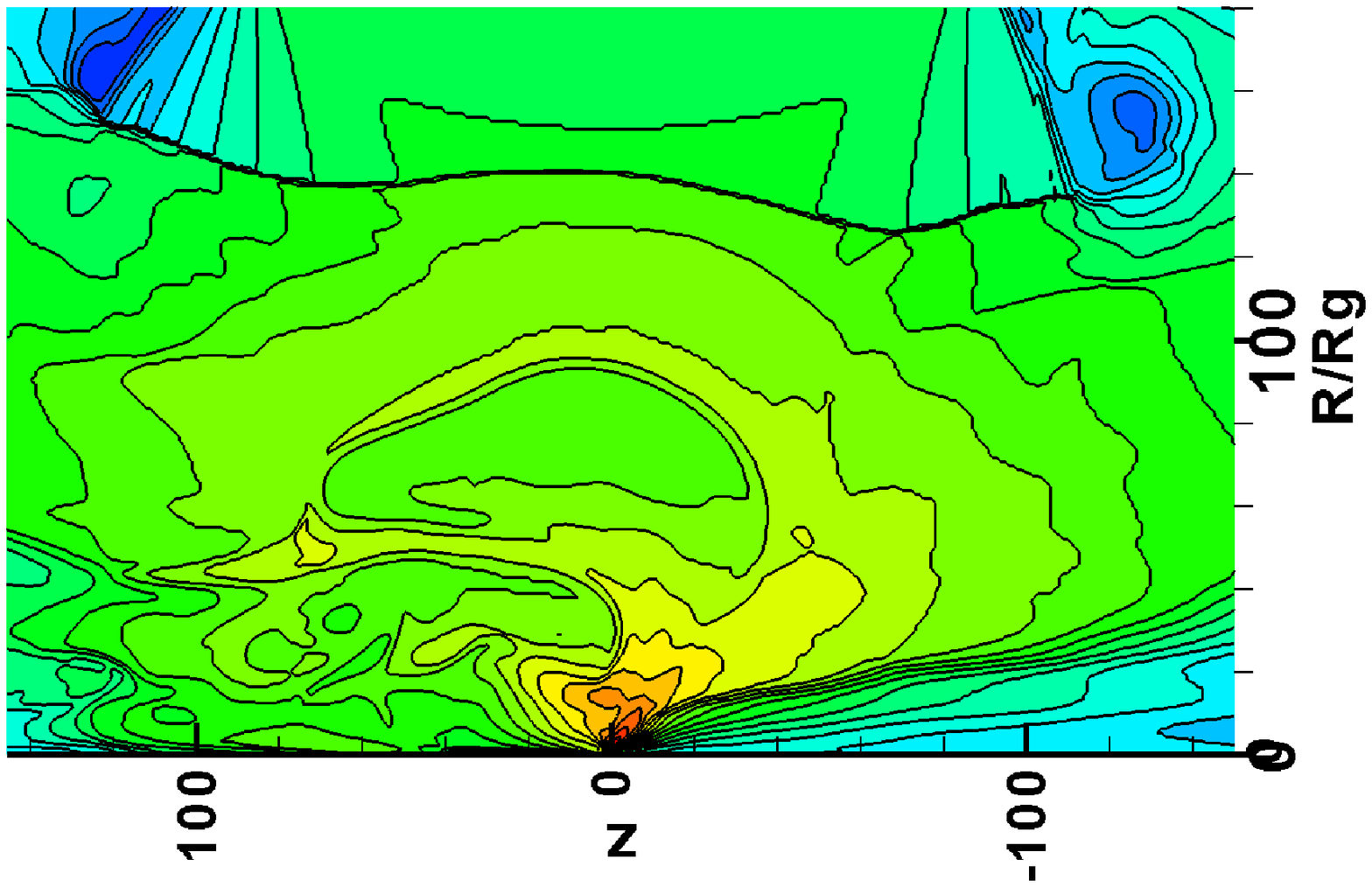}
        \label{ラベル8}
      \end{minipage}

  \begin{minipage}{0.2\linewidth}
        \centering
        \includegraphics[keepaspectratio, scale=0.25,angle=-90]{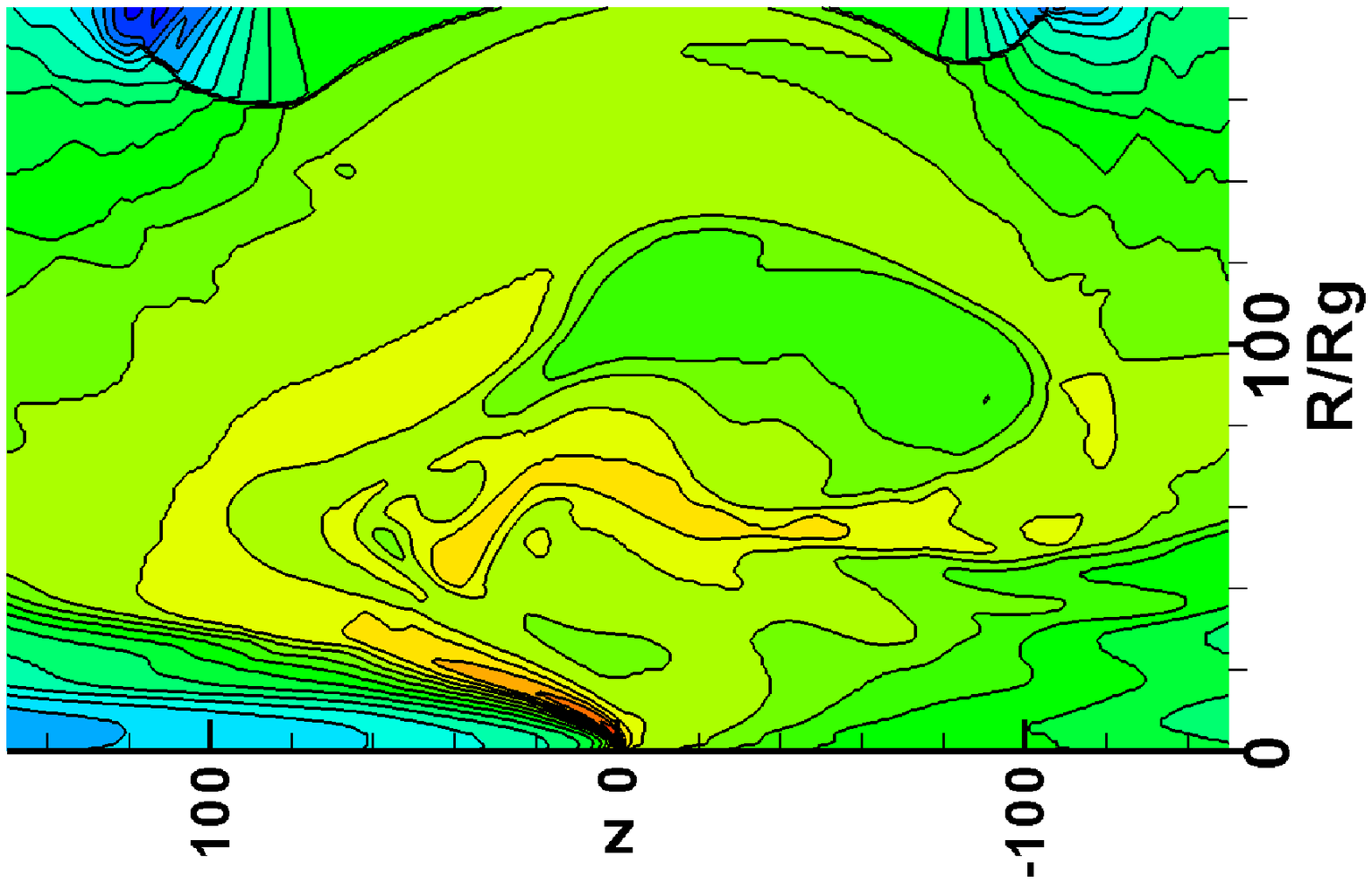}
        \label{ラベル9}
      \end{minipage}
 \begin{minipage}{0.2\linewidth}
        \centering
        \includegraphics[keepaspectratio, scale=0.25,angle=-90]{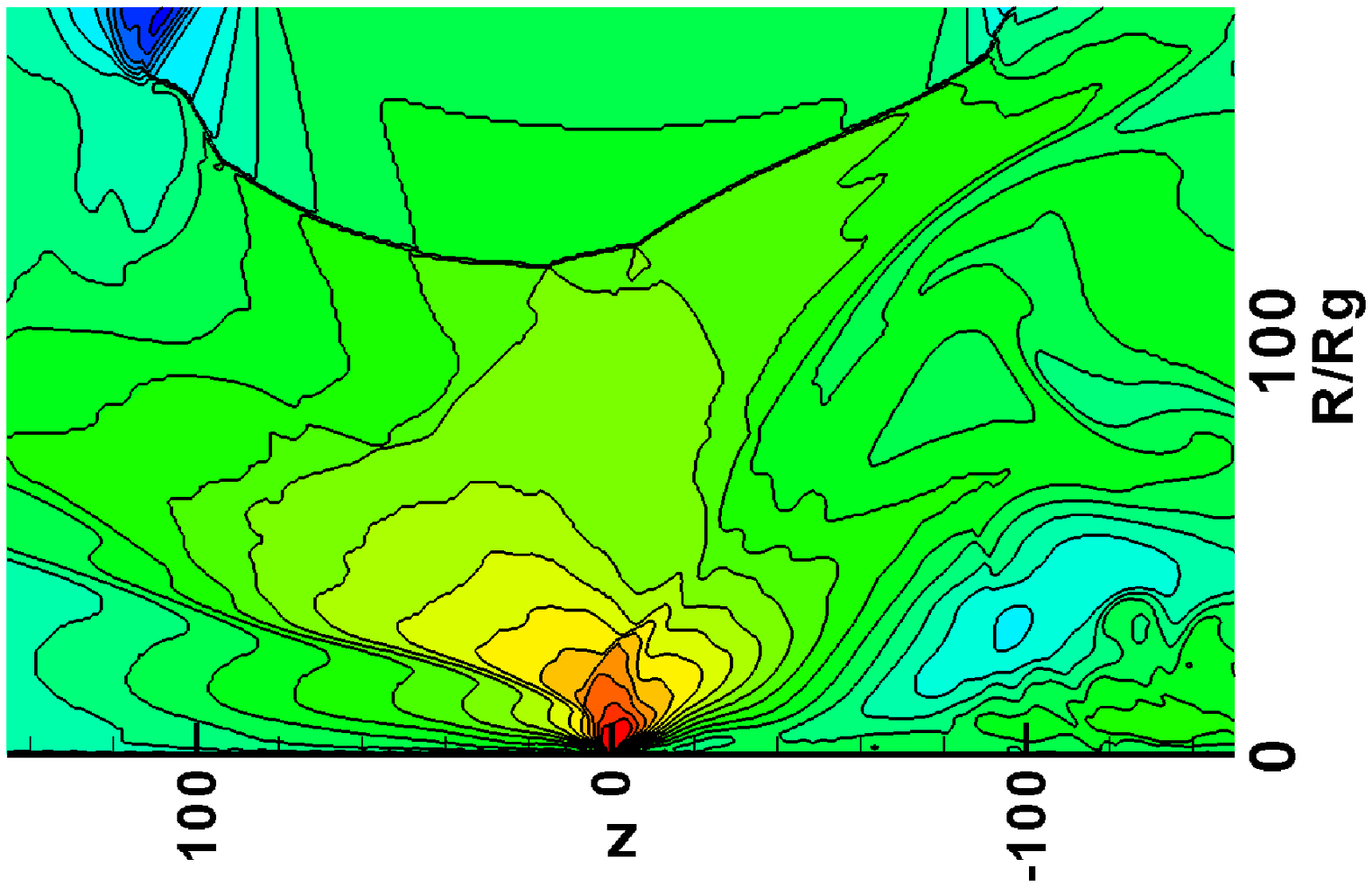}
        \label{ラベル10}
      \end{minipage}\\

      \begin{minipage}{0.2\linewidth}
        \centering
        \includegraphics[keepaspectratio, scale=0.25,angle=-90]{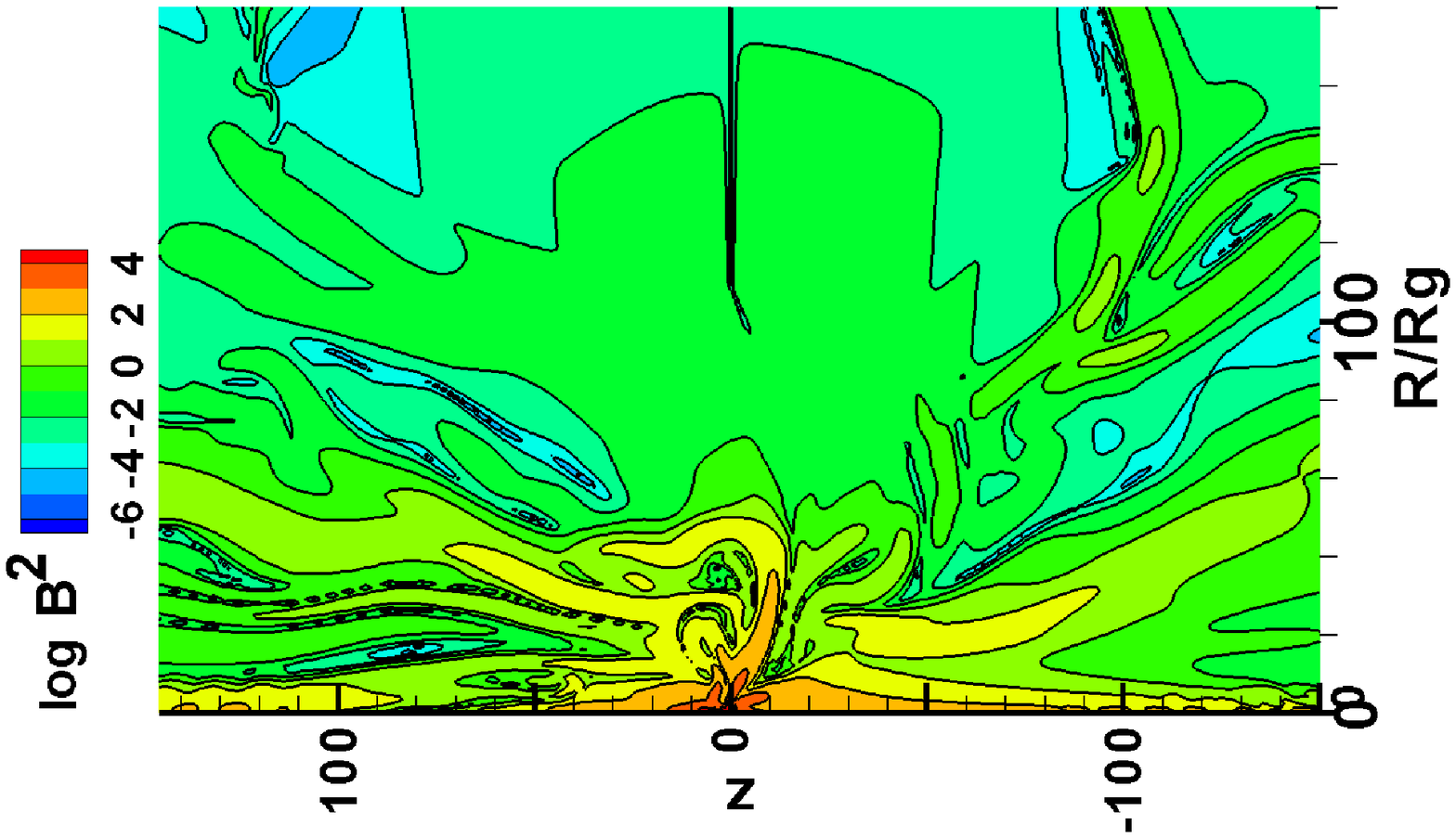}
        \label{ラベル11}
      \end{minipage} 
     
      \begin{minipage}{0.2\linewidth}
        \centering
        \includegraphics[keepaspectratio, scale=0.25,angle=-90]{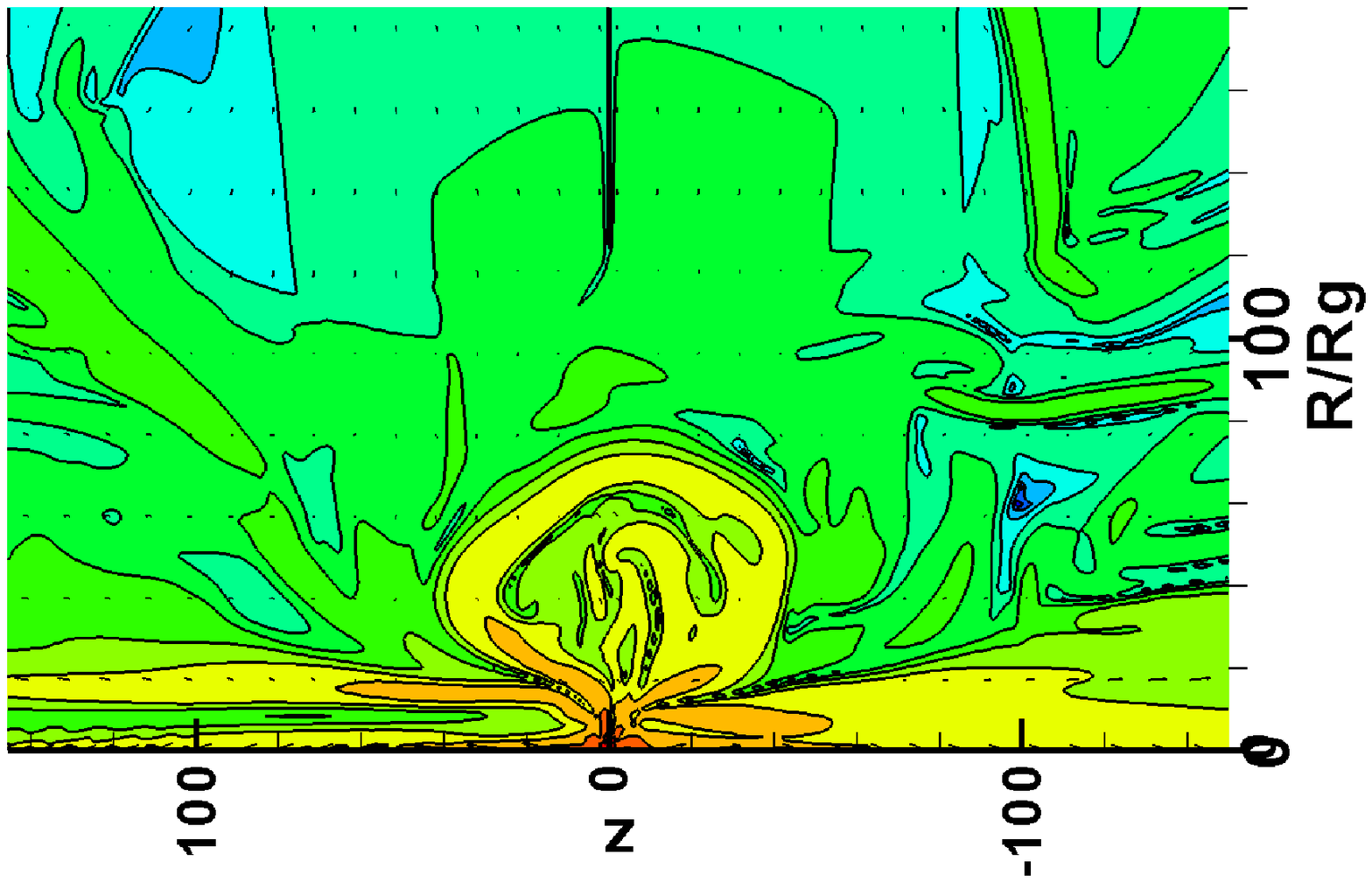}
        \label{ラベル12}
      \end{minipage} 
     
  \begin{minipage}{0.2\linewidth}
        \centering
        \includegraphics[keepaspectratio, scale=0.25,angle=-90]{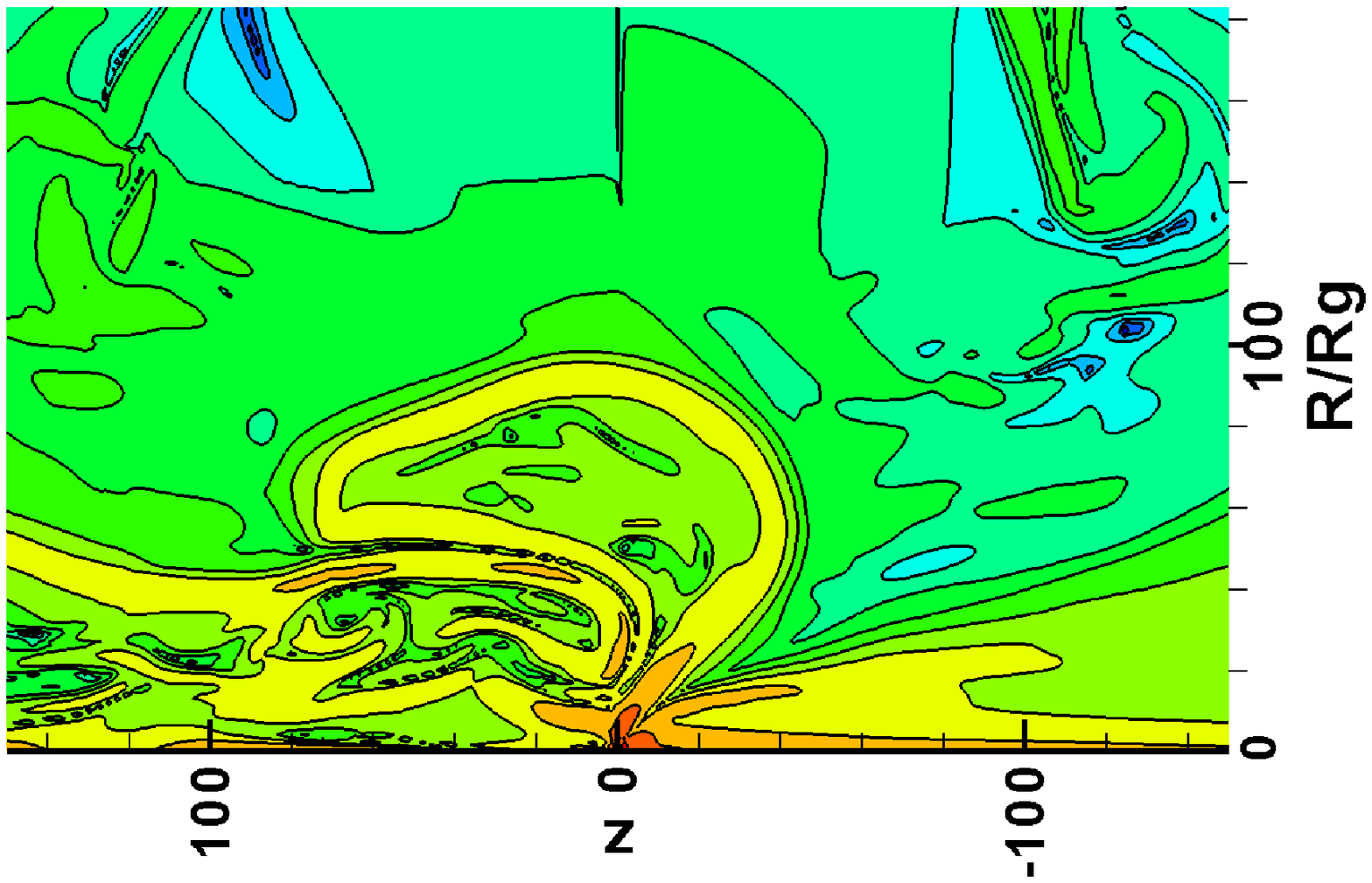}
        \label{ラベル13}
      \end{minipage}
  \begin{minipage}{0.2\linewidth}
        \centering
        \includegraphics[keepaspectratio, scale=0.25,angle=-90]{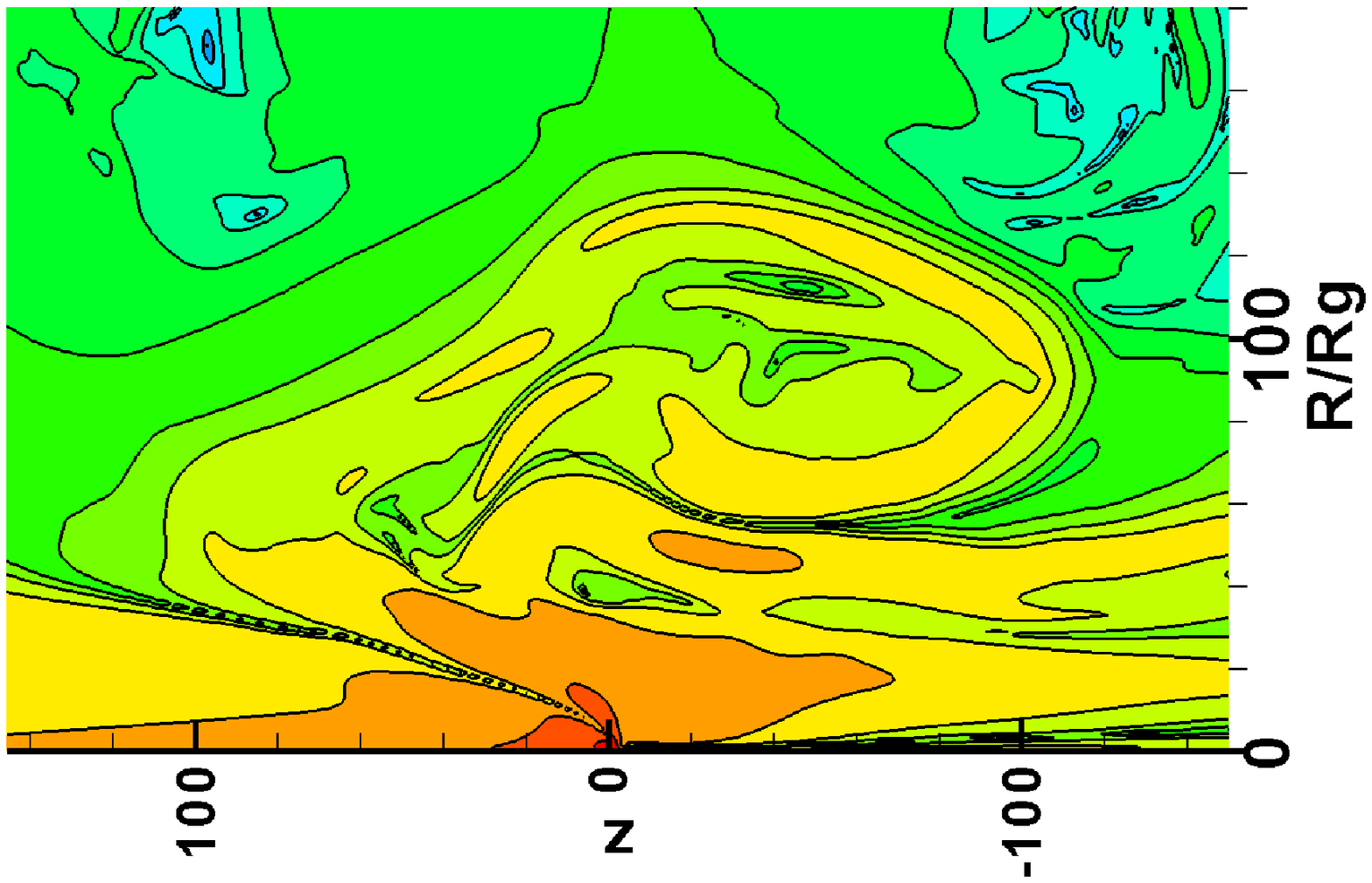}
        \label{ラベル14}
      \end{minipage}

 \begin{minipage}{0.2\linewidth}
        \centering
        \includegraphics[keepaspectratio, scale=0.25,angle=-90]{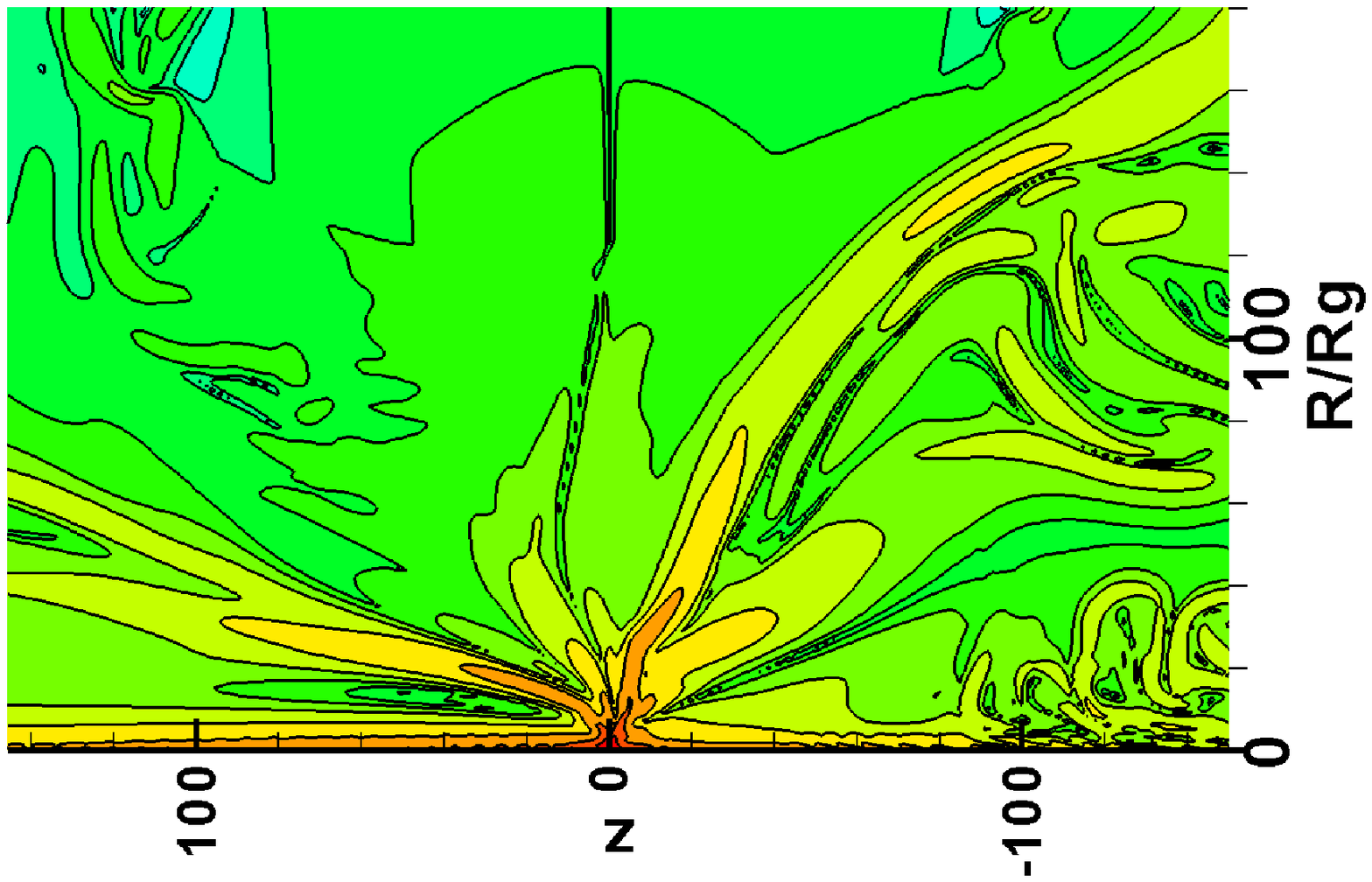}
        \label{ラベル15}
      \end{minipage}\\
      \begin{minipage}{0.2\linewidth}
        \centering
        \includegraphics[keepaspectratio, scale=0.25,angle=-90]{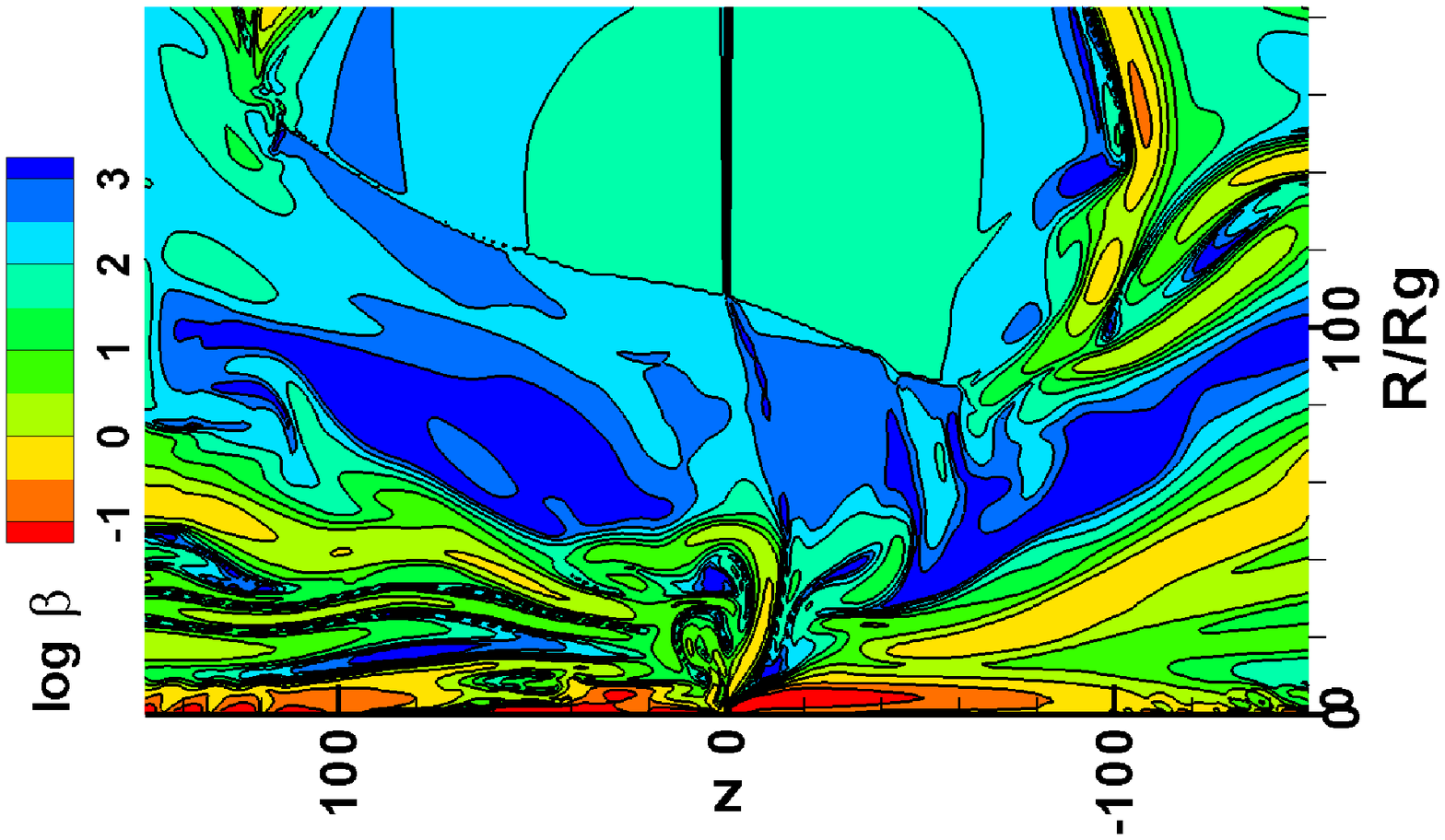}
        \label{ラベル16}
      \end{minipage} 
     
      \begin{minipage}{0.2\linewidth}
        \centering
        \includegraphics[keepaspectratio, scale=0.25,angle=-90]{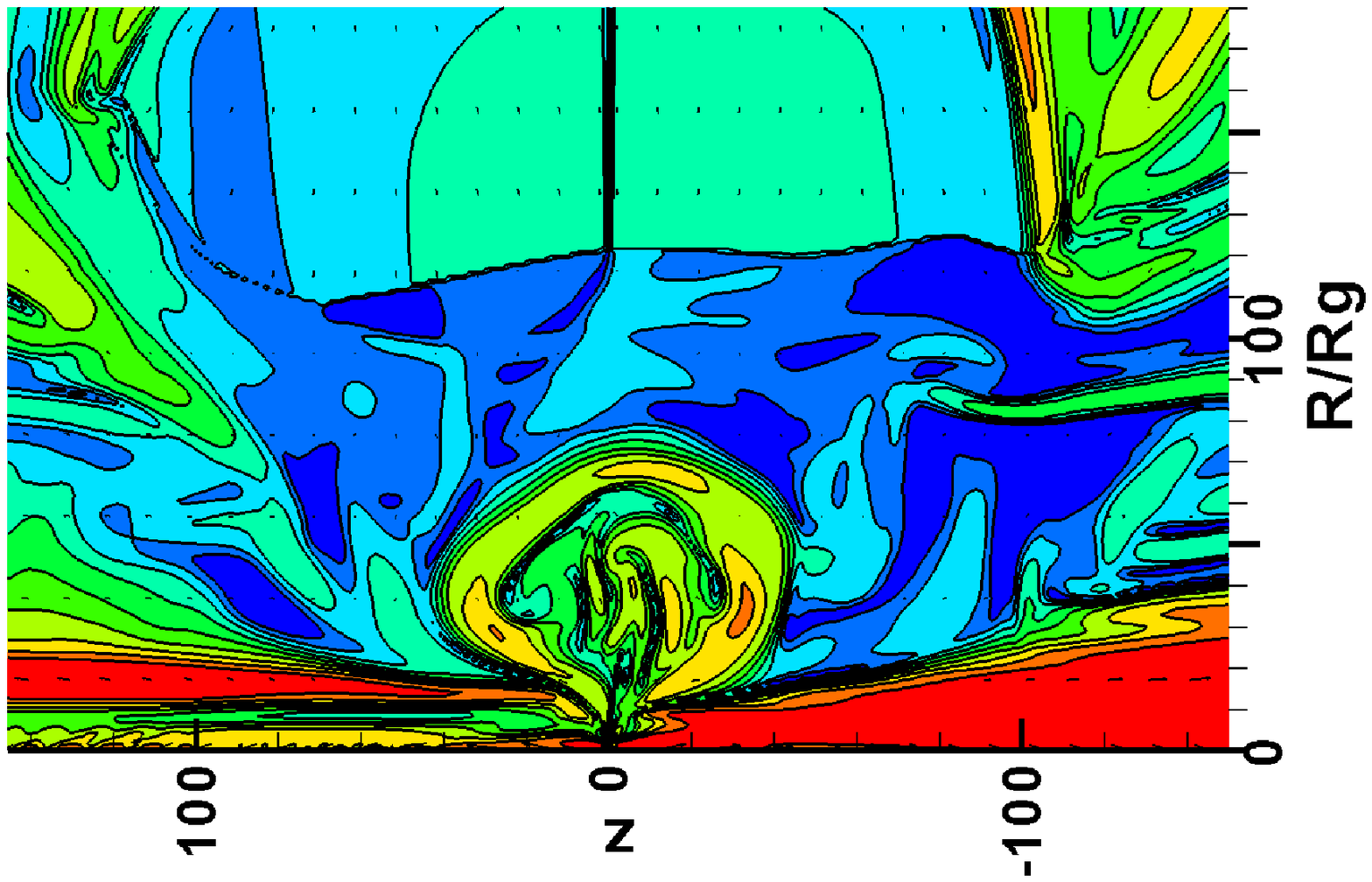}
        \label{ラベル17}
      \end{minipage} 
     
  \begin{minipage}{0.2\linewidth}
        \centering
        \includegraphics[keepaspectratio, scale=0.25,angle=-90]{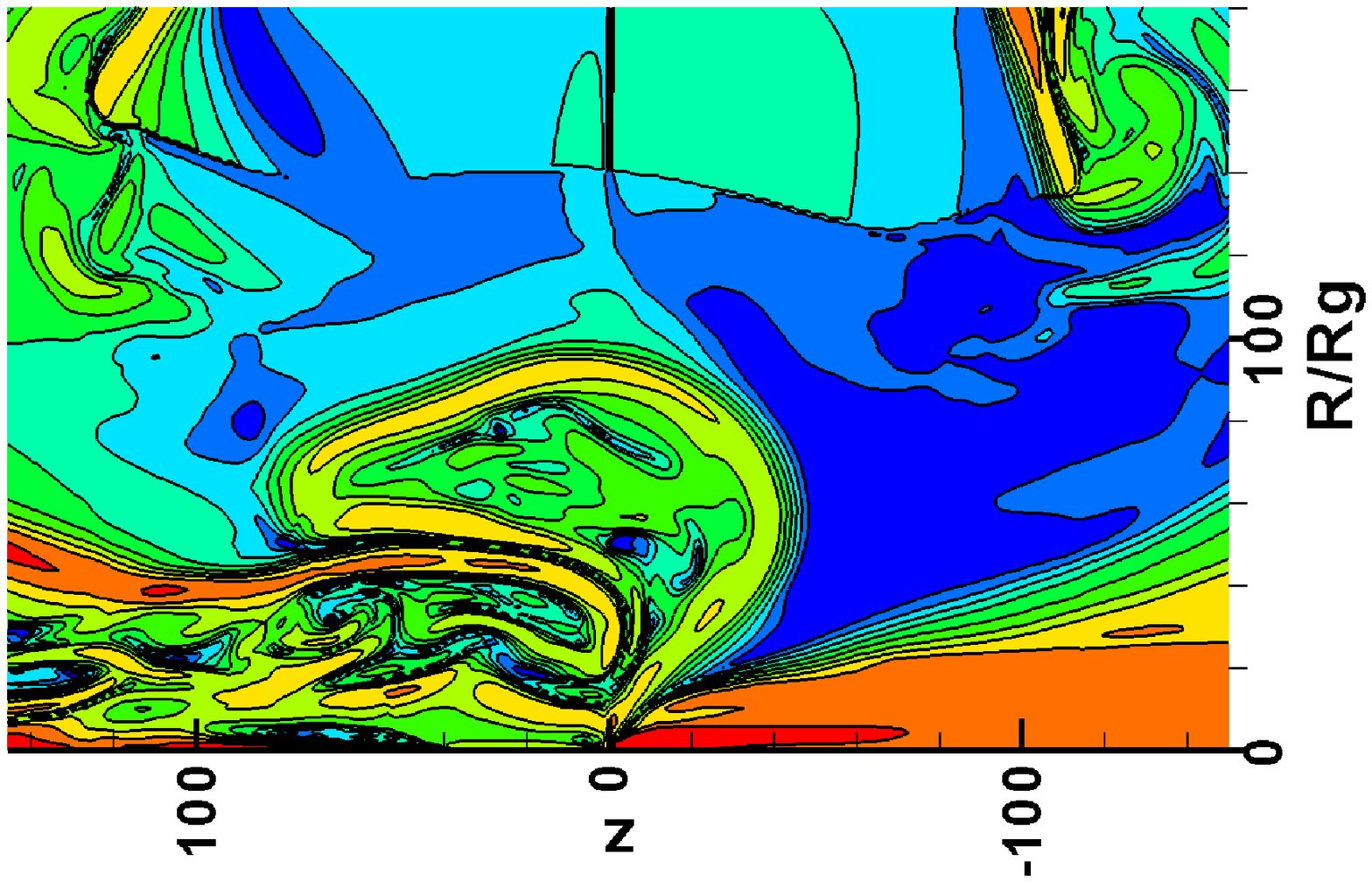}
        \label{ラベル18}
      \end{minipage}
  \begin{minipage}{0.2\linewidth}
        \centering
        \includegraphics[keepaspectratio, scale=0.25,angle=-90]{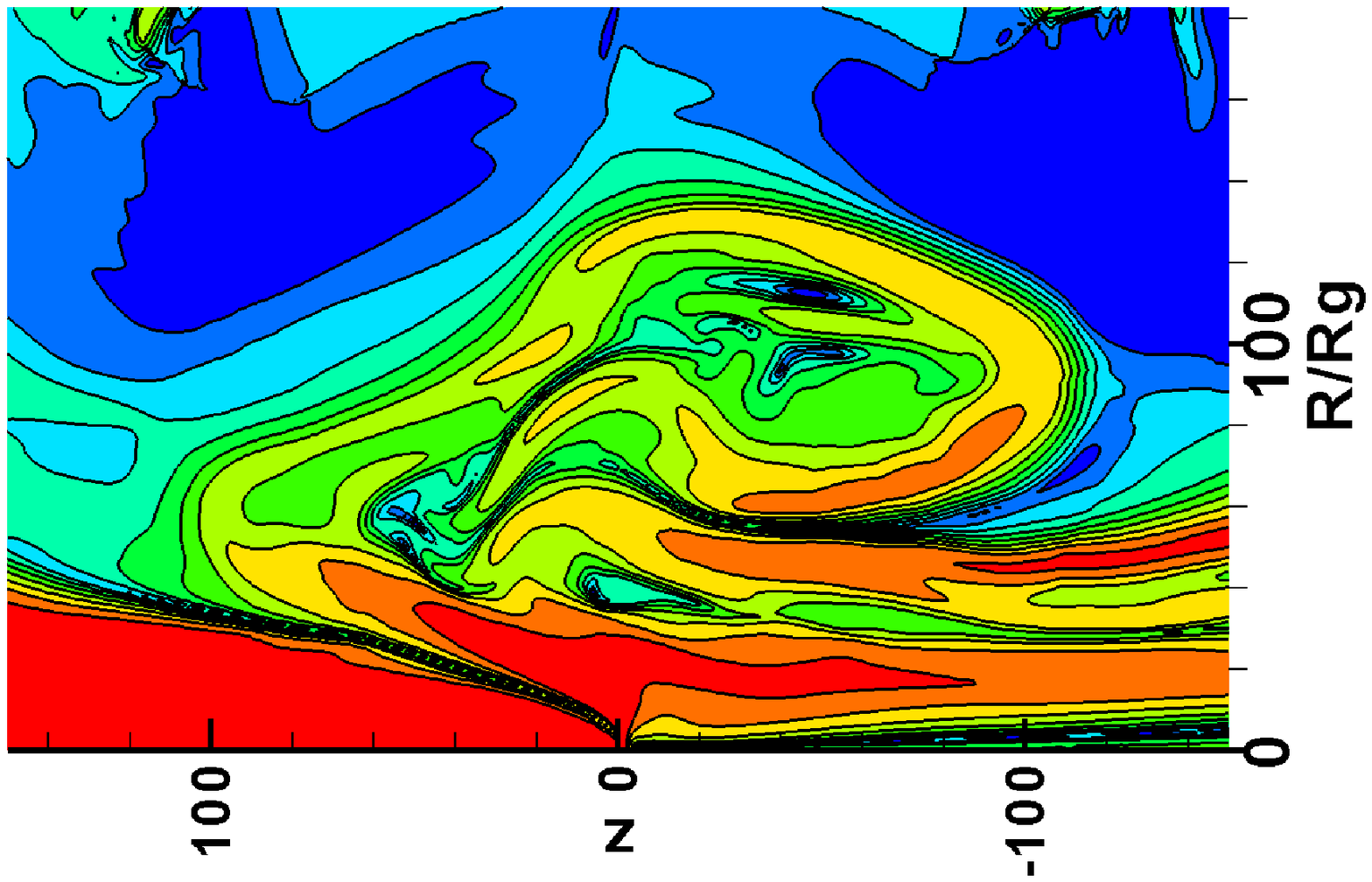}
        \label{ラベル19}
      \end{minipage}

 \begin{minipage}{0.2\linewidth}
        \centering
        \includegraphics[keepaspectratio, scale=0.25,angle=-90]{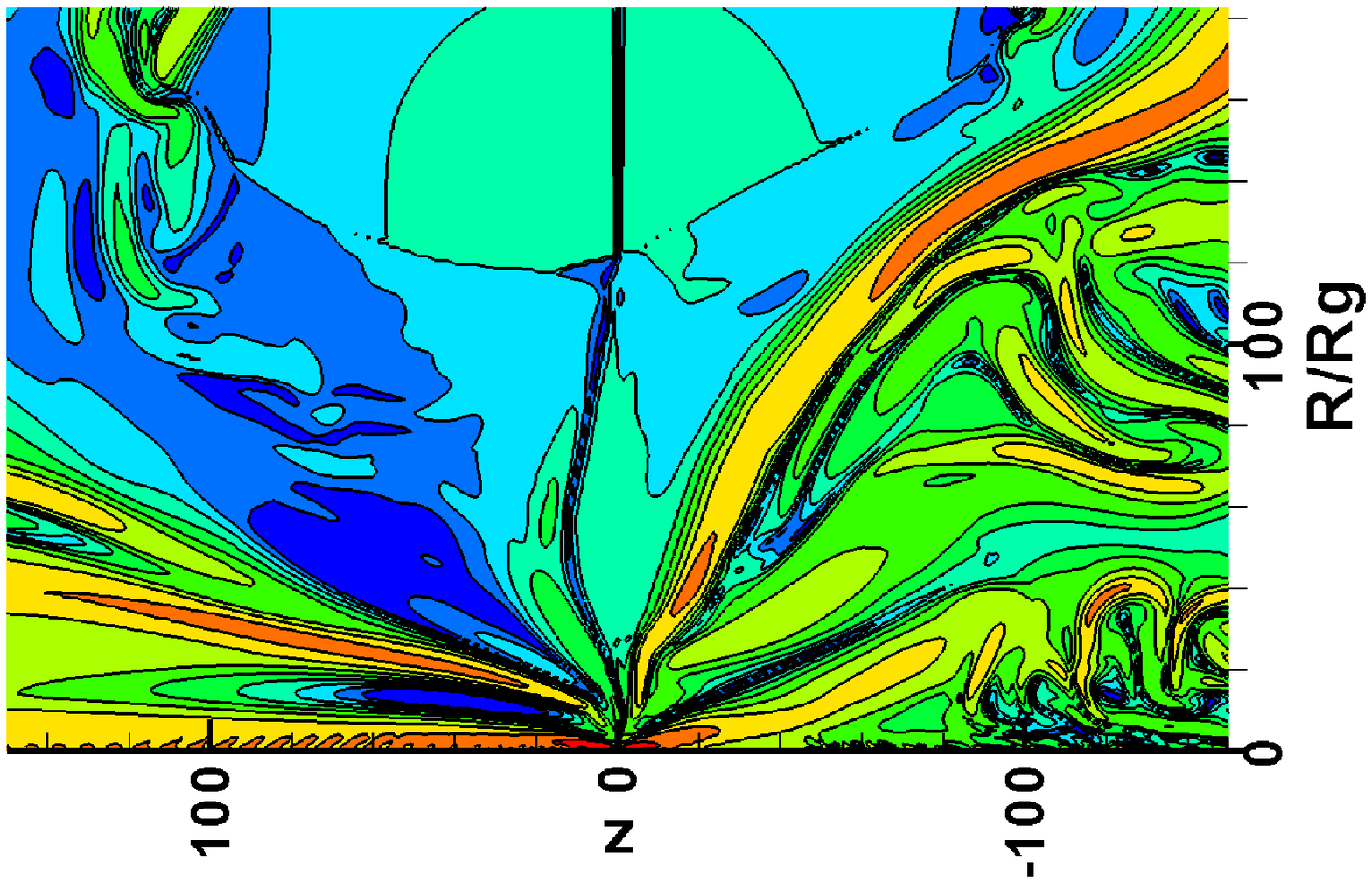}
        \label{ラベル20}
      \end{minipage}
    \end{tabular}
  \vspace*{5mm}
\caption{Time variations of unit velocity vectors and 2D contours of density $\rho$,  magnetic field
  strength $\textbf{B}^2 $ and ratio $\beta$ of  gas pressure to  magnetic pressure 
  at times $ t= 2\times 10^5$ (a), $4\times 10^5$ (b), $8\times 10^5$ (c), $1.2\times 10^6$ (d)
 and $1.6\times 10^6$ (e) s (left to right) in model A.
  The luminosity is minimal in (d) and then becomes maximal in (e), while the shock location at equator is
  maximal and minimal at the former and the latter times, respectively.
  }
  \end{figure}

\subsubsection{Time Variations of the Luminosity and the Shock Location}
We take the output data for the luminosity and the shock location at every time interval of 100 $R_{\rm g}/c$
 ($\sim 4 \times 10^3$ s). Therefore, the time resolution in our simulations is one hour at most.
 Fig.~7 shows the variations of the luminosity $L$  
 and the shock position $R_{\rm s}$  at the equatorial plane in model A with the parameter 
 $\beta_{\rm out}=1000$ of magnetic field strength.
 Here, the arrow at $t=1.6\times 10^6$ s on the abscissa  denotes the epoch (e) during the time evolution
 of the flow described in Fig~5.
The shock and the luminosity oscillate irregularly with  time scales of $\sim 10^5 - 10^6$ s, 
and the average $L$ is $\sim 4.0 \times 10^{34}$  erg s$^{-1}$.
 It should be noticeable that the gap of $R_{\rm s}$ curve at the phase $t \sim 6.5\times 10^6$
  shows a possibility of the moving away of the shock from the outer boundary.
 Fig.~8 shows the time variations of the mass-outflow rate $\dot M_{\rm out}$ and 
 the mass-inflow rate $\dot M_{\rm edge}$ at the inner edge in model A. 
 Here, in order to compare more clearly the phases of their variations, we take different vertical scales 
 for $\dot M_{\rm out}$ and $\dot M_{\rm edge}$. 
  $L$, $R_{\rm s}$, $\dot M_{\rm out}$ and  $\dot M_{\rm edge}$ in model A  show  irregular, but recurrent 
 variations roughly in a time-scale of $\sim 5 \times 10^5$ s. 
These irregular oscillations remain constant without fading.
 From the time variations of  $L$, $R_{\rm s}$, $\dot M_{\rm out}$ and  $\dot M_{\rm edge}$,
 we find strong correlations between $L$ and  $\dot M_{\rm edge}$ and between  $R_{\rm s}$
 and $\dot M_{\rm out}$ and an anti-correlation between $L$ and  $R_{\rm s}$.
 
 \begin{figure}
 \begin{center}
  \begin{tabular}{c}
  \begin{minipage} {0.5\linewidth}
   \begin{center}
    \includegraphics[width=80mm,height=70mm,angle=0]{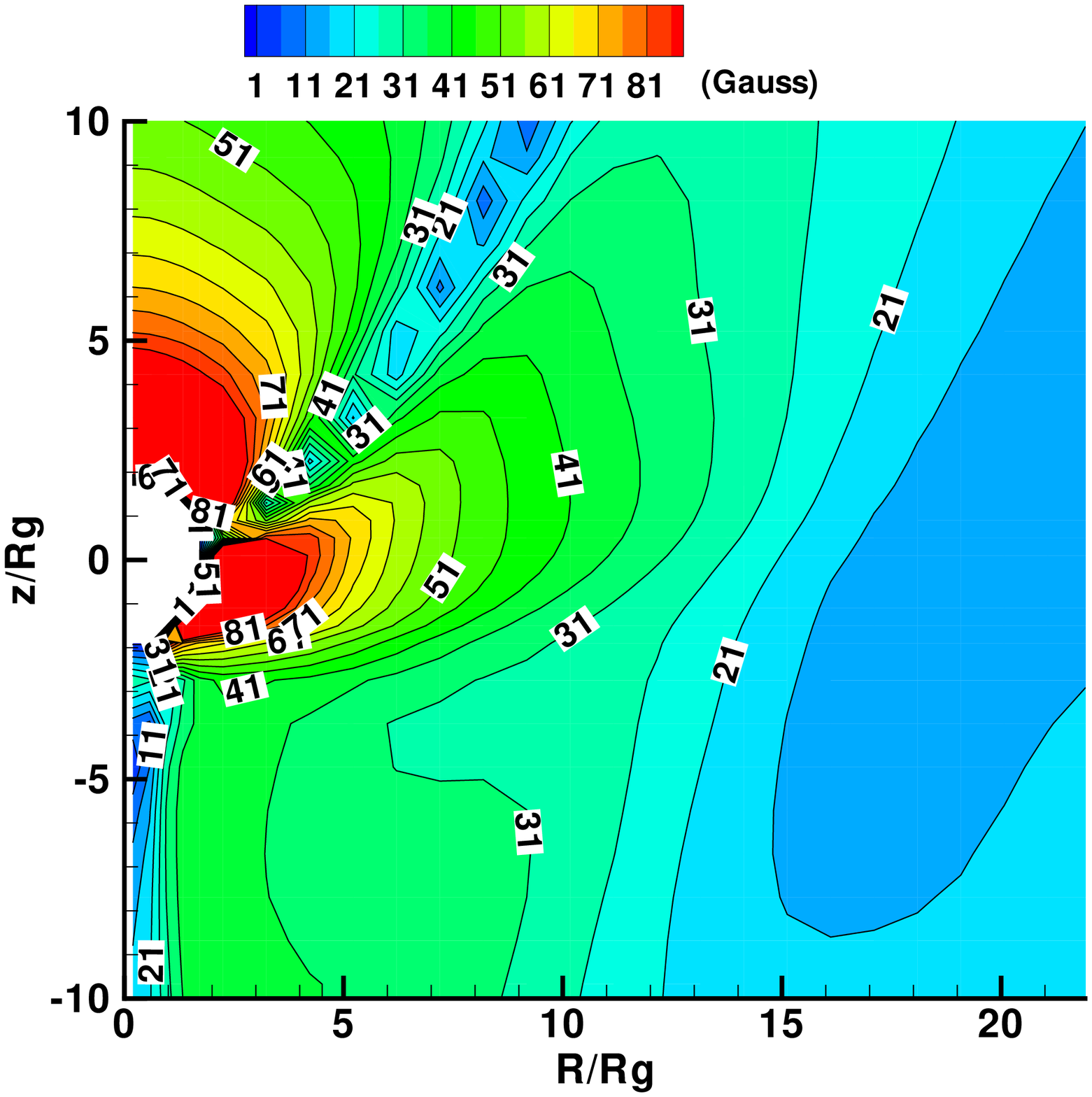}
     \label{fig12-1}
 \end{center}
 \end{minipage}

 \begin{minipage}{0.5\linewidth}
  \begin{center}
   \includegraphics[width=80mm,height=70mm,angle=0]{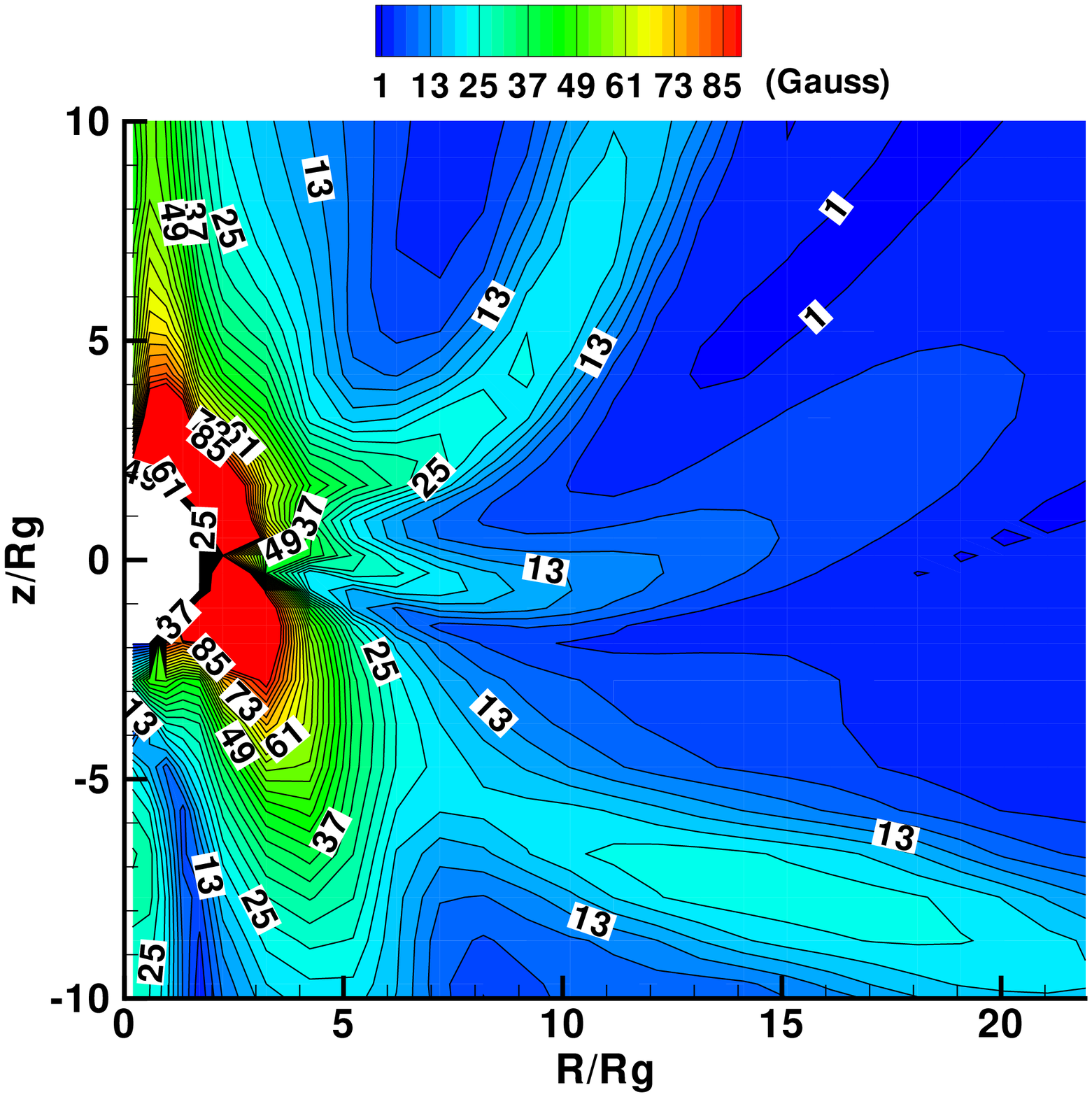}
     \label{fig12-2}
  \end{center}
  \end{minipage}
  \end{tabular}
 \vspace*{5mm}
 \caption{2D contours of the strength  $\mid \textbf{B} \mid$ (Gauss) of the magnetic field in the inner region at
  $t = 1.2\times 10^6$ s (left) and $1.6\times 10^6$ s (right) for model A. 
 Two distorted central masses of high magnetic field in each contour are very unstable and yield filamentous 
 projections of the magnetic field which develop into spherical bubble-like shape in the outer region.
}
 \end{center}
 \end{figure}

 The overall evolution of the flow  is similar to the initial one during 
 $t \leq 2\times 10^6$ s described in sub-subsection 3.2.2.
 The shock location $R_{\rm s}$ at the equator is initially at $R_{\rm s} \sim 65R_{\rm g}$.
The shock and the high magnetic blob within the PSC begin to expand with increasing  magnetic pressure
 and the shock reaches a maximal location $R_{\rm s} \sim$ 187$R_{\rm g}$ at 
$t= 1.2\times 10^6$ s (phase (d) of Fig. 5) and then recedes back,  while the expanding
high magnetic blob is diffused out. 
 When the shock is expanding through the outermost region, a new high density blob with high temperature 
appears in the inner region  and another inner shock is formed in front of the expanding high density blob, 
 due to the interaction of the blob with the accreting matter.
During the successive evolution, the outer and inner shocks show complex behavior  and
these processes are repeated irregularly.
When the outer shock expands to its maximal position, the mass-outflow rate is maximal and the luminosity
 attains its minimal value. Conversely, when the shock shrinks to its minimal position, the luminosity
  attains the maximal values, while the mass-outflow rate is minimal.
 This behavior is similar to the relation between the luminosity and the pulsating radius in
  variable stars as well known \citep{key-11}.
 We find also that most of the luminosity is emitted from the PSC region and the contribution
 from the outside of the PSC region is less than 10 percent of the total luminosity.
 After the MRI activity settles to a stable state,  the shock finally moves back and forth between 
 $R_{\rm s}$= 60 -- 170$R_{\rm g}$ with an irregular time interval of $\sim 5 \times 10^5$ s. 
Due to the variable shock location, the luminosity varies by more than a factor of 3.
Fig.~9 shows the time variations of $L$  and  $R_{\rm s}$  for model B with the parameter 
 $\beta_{\rm out}=5000$ of magnetic field strength.
The shock and the luminosity oscillate irregularly with  time scales of $\sim 10^5 - 10^6$ s 
and the average $L$ is $\sim 3.0 \times 10^{34}$  erg s$^{-1}$, i.e. the same as model A.
  The maximum shock locations in model B are a little smaller than those in model A, 
 because of its smaller magnetic field strength.

 Fig.~10 shows the power density spectra of luminosity $L$ for models A (Left) and B (Right). 
 Though we can not find very clear peaks of the power density spectra in both models,
 model A shows a break frequency at  $ 2.0\times 10^{-6}$ Hz together with another weak signal
  at $\sim 1.1\times 10^{-5}$ Hz. On the other hand,  model B denotes peak frequencies at
 $2.5 \times 10^{-6}$ and $1.1 \times 10^{-5}$ Hz, including two more additional  weak features 
 at $5.4 \times  10^{-6}$ and $3.0 \times 10^{-5}$ Hz. These frequencies may be harmonics of the original 
 oscillation with  $ 2.5 \times 10^{-6}$ Hz.
 As a result, we observe a peak signal at frequency (2.0 -- 2.5) $ \times 10^{-6}$ Hz along with
  a common weak signature at $1.1\times 10^{-5}$ Hz in both models, 
 which correspond to  periods  (4.6 -- 5.8) ($\sim$ 5) days and 25 hrs, respectively.
 The larger period $\sim 5$ days corresponds to the time-scale of the irregular shock oscillation 
 between 60$R_{\rm g} \leq R \leq 170R_{\rm g}$. 
 Fig.~11 shows the Mach number of the radial velocity on the equator at times
 $t =  4.6 \times 10^6$ (solid line),  $5.7 \times 10^6$ (dash-dot line) and 6.2 $\times 10^6$ (dashed line) s in model A.  
 Here, there exist another shock phenomena behind the outer shock for each curve.
  The shock corresponds to the inner shock mentioned in sub-subsection 3.2.3.
 The inner shocks are weak compared with the outer shock and 
 the shock features are complicated.  From the animation of  Mach number 
 versus radius, we  recognize that the inner shock oscillates irregularly and rapidly.
 The inner shocks also contribute to the luminosity because the density and the temperature behind the 
 inner shock are higher than those behind the outer shock, although the Mach number is smaller
 than that in the outer shock.
 However, the variability pattern of the inner shock is not clear as the one found in the outer shock oscillation 
 and may be recognized as a weak signature at  $\nu$ = 1.1$\times 10^{-5}$ Hz in the power density spectrum.
 We conclude that the time-variability with two different periods in models A and B is due
 to an oscillating outer strong shock  with another more rapid oscillation of the inner weak shock.

\begin{figure}
\begin{center}
\includegraphics[width=0.8\textwidth]{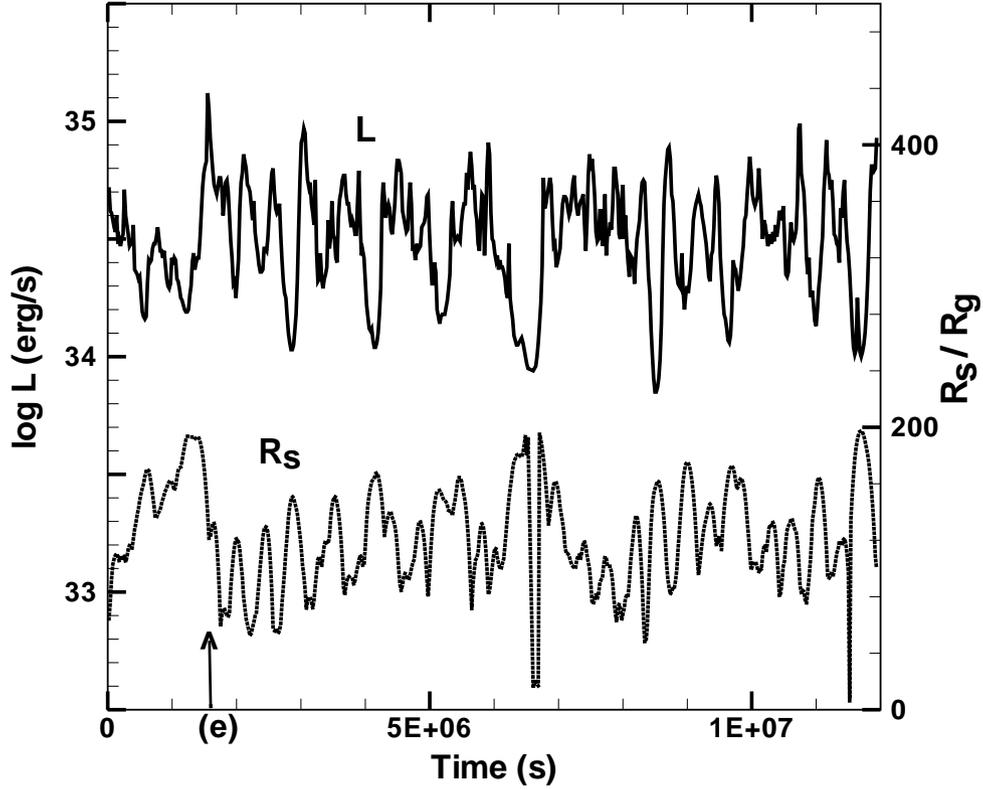}
\end{center}
\caption{Time evolutions of luminosity $L$ (erg s$^{-1}$)  and 
     shock position $R_{\rm s}$ (in $R_{\rm g}$ units)  at the equatorial plane in the magnetized model A
 with the parameter  $\beta_{\rm out}=1000$ of magnetic field strength.
  When the shock expands to a maximal position, the luminosity attains its minimal value.
  Conversely, when the shock shrinks to its minimal position, the luminosity is maximal.
  The arrow at $t=1.6\times 10^6$ s on the abscissa  denotes the epoch (e) during the time evolution
 of the flow described in Fig~5.
}
\end{figure}

\begin{figure}
\begin{center}
\includegraphics[width=0.7\textwidth]{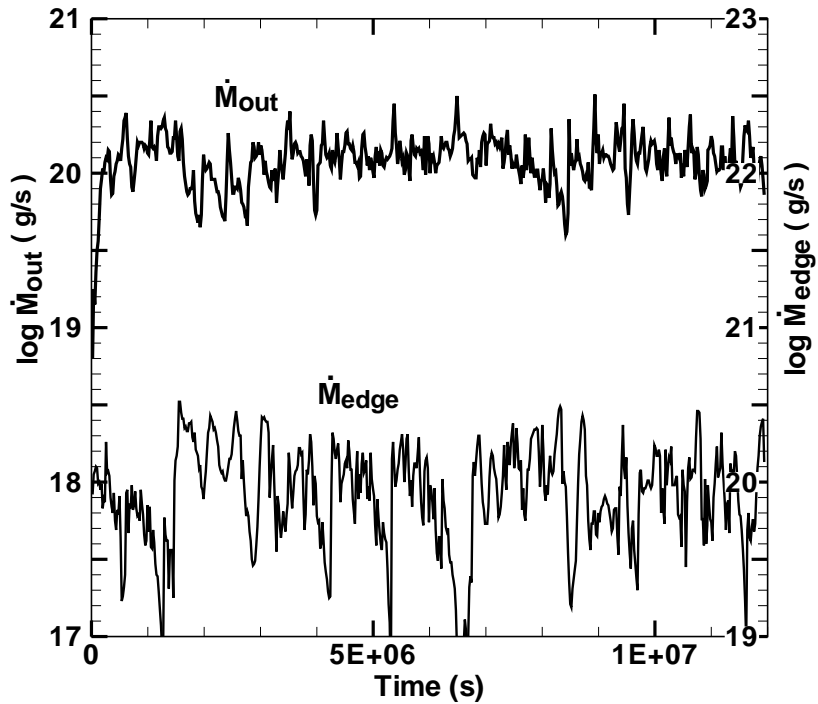}
\end{center}
\caption{
Time evolutions of  the mass-outflow rate $\dot M_{\rm out}$ (g s$^{-1}$) and
 the mass-inflow rate $\dot M_{\rm edge}$ (g s$^{-1}$)  at the inner edge in model A.
 To compare clearly the phases of their variations, different vertical scales for $\dot M_{\rm out}$ 
 and $\dot M_{\rm edge}$ are taken here.
. 
}
\end{figure}

\begin{figure}
\begin{center}
\includegraphics[width=0.7\textwidth]{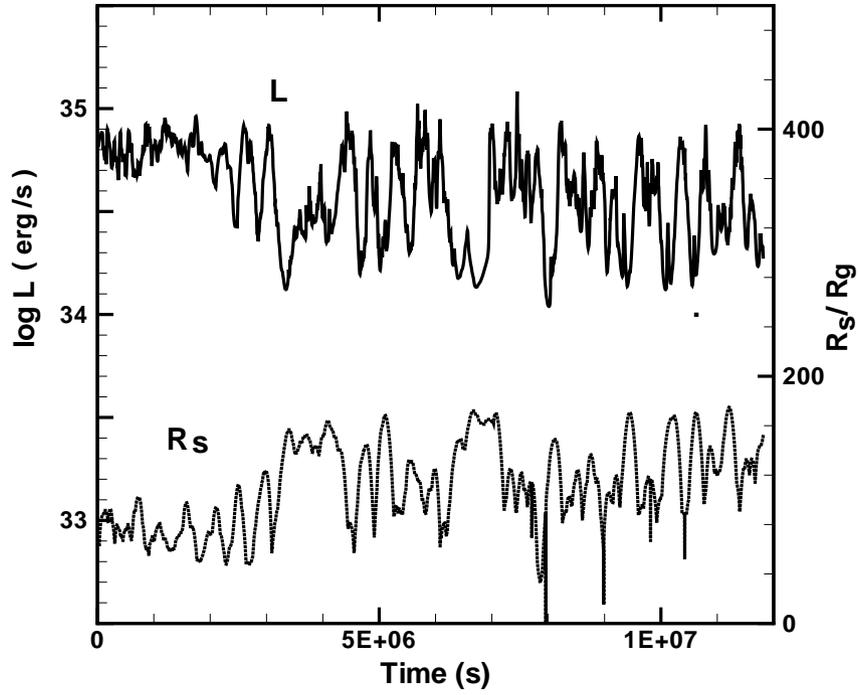}
\end{center}
\caption{
 Same as Fig~7 but for model B with the parameter $\beta_{\rm out}$ = 5000 of 
 magnetic field strength.
}
\end{figure}

\begin{figure}
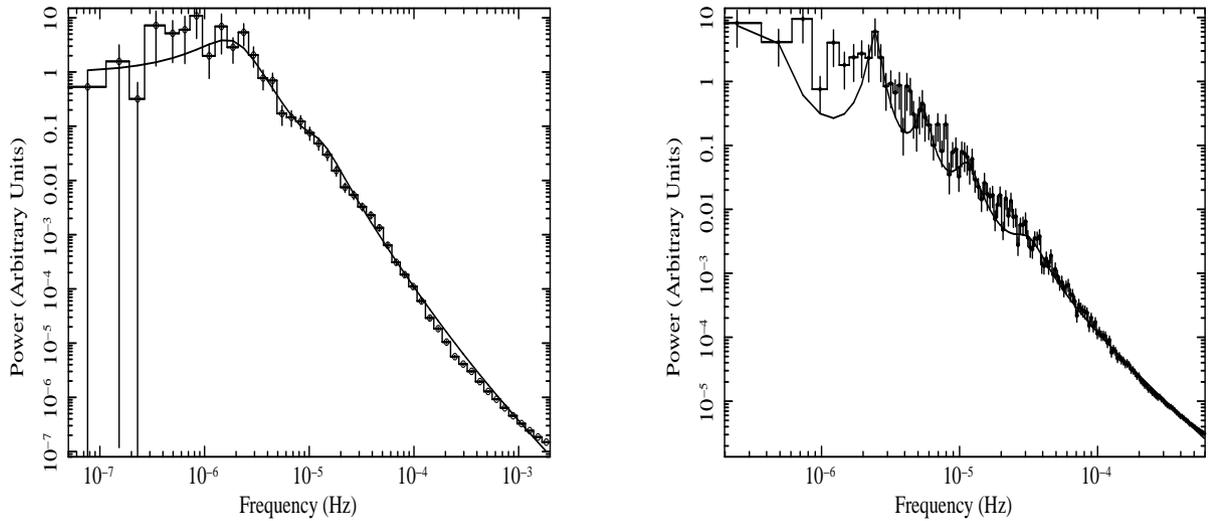

 \begin{center}
  \begin{tabular}{c}
  \begin{minipage} {0.5\linewidth}
   \begin{center}
    \includegraphics[width=70mm,height=80mm,angle=270]{fig10-1.eps}
     \label{fig12-1}
 \end{center}
 \end{minipage}

 \begin{minipage}{0.5\linewidth}
  \begin{center}
   \includegraphics[width=70mm,height=80mm,angle=270]{fig10-2.eps}
     \label{fig12-2}
  \end{center}
  \end{minipage}
  \end{tabular}
 \caption{Power density spectra of luminosity $L$ for models A (left)
 and  B (right). A break frequency at  $ 2.0\times 10^{-6}$ Hz  with another weak signal
  at $\sim 1.1\times 10^{-5}$ Hz in model A and peak frequencies at
 $2.5 \times 10^{-6}$ and $1.1 \times 10^{-5}$ Hz in model B are observed.
}
 \end{center}
 \end{figure}

\begin{figure}
\begin{center}
\includegraphics[width=100mm,height=80mm]{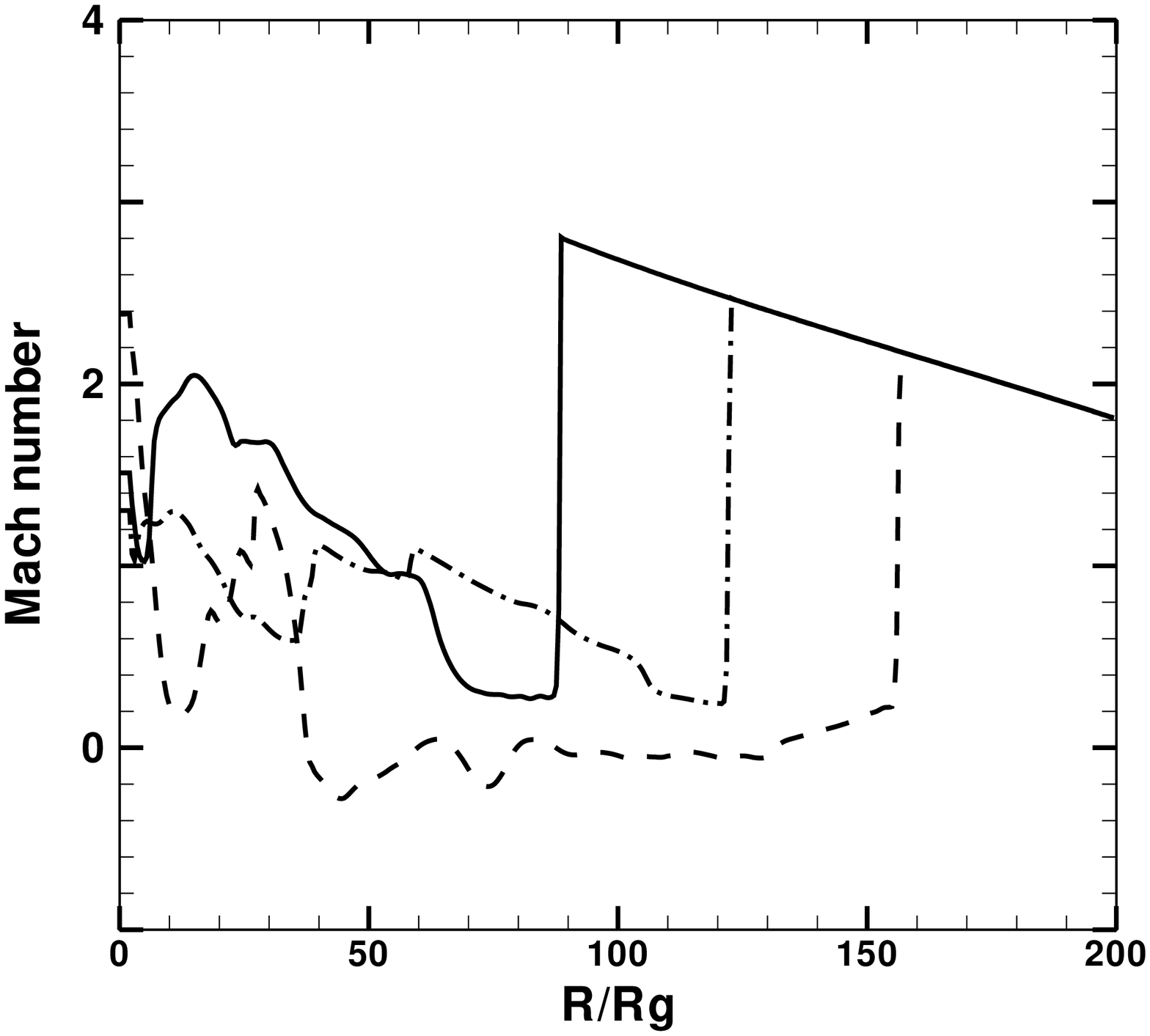}
\end{center}
\caption{
 Radial profiles of Mach number of the radial velocity -$v_{\rm R}$ on the equator at $t= 
  4.6 \times 10^6$ (solid line),  $5.7 \times 10^6$ (dash-dot line) and $6.2 \times 10^6$ (dashed line)
 s in model A. The outer strong shock oscillates at 60 $\leq R \leq$ 170
 and the inner weak shock oscillates irregularly and rapidly
 far below the outer shock.}
\end{figure}

 Our magnetized flow are compared with the same as reported by \citet{key-51-2}.
 They started with the initial conditions of the Bondi spherical accretion flow, but with a constant angular momentum
  $\lambda$ in the region of $- h_{\rm out} \leq z \leq h_{\rm out}$ at the outer boundary
  and used values of  $\lambda$ = 1 -- 2 similar to ours, but Bernoulli constant $\epsilon=0$,  instead of 
 our positive value as $\epsilon =1.98 \times 10^{-6}$.
  The 1.5D transonic solutions for these parameters of $\epsilon$ and $\lambda$ never yield the standing 
  shock formation under the vertical hydrostatic equilibrium assumption.
  They examined the magnetized flow over a wide range up to the Bondi radius and with $\beta_{\rm out}$ =
   $10^5$ -- $10^7$.  Their results of the global MRI effects on the angular velocity, the magnetic pressure
   and the Maxwell stress and of the time variation of the accretion rate are similar to those in our models.
   But the quasi-periodic oscillation (QPO)-like features of the accretion rate are not found in their results.
   Because the QPO-like variabilities of the luminosity and the mass outflow rate
   in our models are due to the shock wave oscillation driven by the magnetic field.

\subsubsection{Relevance to Sgr A*}
The supermassive black hole candidate Sgr A* shows quiescent and flare states.
The rapid flares over hours to day of Sgr A* have been detected in multiple wavebands from radio, 
 sub-millimetre and IR to X-ray. 
 Recent Chandra, Swift and XMM-Newton observations over long durations show that flares
 with X-ray luminosity $L_{\rm x} > 10^{34}$ erg s$^{-1}$ occur at a rate of $\sim1$ per day, 
 while luminous flares with $L_{\rm x} > 10^{35}$ erg s$^{-1}$ occur  every  5 -- 10 days
  (Degenaar et al. 2013; Neilsen et al. 2013, 2015; Ponti et al. 2015).
 The results of the time dependent simulations in this work may be compared with these 
 long-term flares of  Sgr A*.
 In the present magnetized models A and B, the luminosities vary as 
 $8 \times 10^{33}$ -- $1.4 \times 10^{35}$ erg s$^{-1}$. 
 The averaged maximal luminosity due to the outer shock is $\sim 1.0 \times 10^{35}$ erg s$^{-1}$ and 
 the variable maximal luminosity due to the inner shock is roughly  estimated to be $\sim 2.5 \times 10^{34}$
 erg s$^{-1}$.

 However, it should be noticed that the above luminosities might be underestimated when the synchrotron radiation
  dominates the free-free emission as mentioned in sub-subsection 3.2.1.
 \citet{key-65} proposed reasonable models of the quiescent and rapid flare states in X-ray spectra,
 considering synchrotron and Inverse Compton emissions by accelerated electrons, where the X-ray radiation
 emitted at the flare state is two orders of magnitude larger than that at the quiescent state.
  They showed the quiescent model with a strong magnetic field of  $\mid \textbf{B} \mid \sim$ 20 G 
 at $R \leq 10R_{\rm g}$ 
 and the X-ray flare model with a weak magnetic field $\mid \textbf{B} \mid \leq$ 1 G in the emitting region.
 From the best fit parameters of the mean spectrum of very bright flares, 
 \citet{key-50}  showed that large magnetic field amplitude ($\mid \textbf{B} \mid \sim$ 30 G) 
 is observed at the start
 of the X-ray flare and then drops to $\mid \textbf{B} \mid \sim 4.8$ G at the peak of the X-ray flare.
 This scenario to the rapid flares  may be adaptable also to the long-term flares of Sgr A*.
 In this respect, it is useful for us to refer the time evolution of the magnetic field strength in Fig~5. 
 The luminosity is minimal as $\sim 1.6 \times 10^{34}$ erg s$^{-1}$  at $t=1.2\times 10^6$ s (d)
 and then attains the maximal of $\sim 1.1 \times 10^{35}$ erg s$^{-1}$ at $1.6 \times 10^6$ s (e). 
 This time sequence is regarded as the evolution of the quiescent state to flare state.
 The analyses of the magnetic field evolution show that $ \mid \textbf{B} \mid \geq$ 30 G in (d) and $ \mid \textbf{B} \mid \geq$ 7 G in (e) at $R \leq 10R_{\rm g}$ on the equator as mentioned in sub-subsection 3.2.2. 
 That is, the features of the magnetic field strength at the inner region of the flow in (d) and (e) 
 qualitatively represent well those at the quiescent and X-ray flare states, respectively, shown by 
 \citet{key-65} and \citet{key-50}.
 This suggests strongly that actual maximal luminosity in our models  may increases considerably
 through the synchrotron radiation in the inner region and we expect the maximal luminosity
  due to the outer shock exceeds far $\sim 10^{35}$  erg s$^{-1}$.
 On the other hand, an upper limit on the quiescent luminosity of Sgr A* above 10 keV
 was recently derived to be  $L_{\rm X_{\rm q,10-79 keV}}$  $\leq 2.9\times 10^{34}$ erg s$^{-1}$ \citep{key-72}.
 If we regard the minimal luminosity $\sim 10^{34}$ erg s$^{-1}$ in our models as the quiescent luminosity,
  the luminosity variation of  $L > 10^{34}$ erg s$^{-1}$ occurring at a rate of  25 hrs and 
  $L > 10^{35}$ erg s$^{-1}$ occurring approximately every  5 days in this work is qualitatively 
 compatible with the observed long-term flares of Sgr A*.

\section{Summary and Discussion}
  We examined the effects of magnetic field on low angular momentum flows with
 standing shock around the  black hole and,  adopting fiducial parameters of a specific energy
 $\epsilon$ = $1.96 \times 10^{-6}$, a specific angular momentum $\lambda = 1.35$, 
 and magnetic parameters $\beta_{\rm out}$= 1000 (model A) and 5000 (model B),  applied the magnetized 
 flow models to the long-term flares of Sgr A*.
 The results  of our 2D simulations are summarized below.

 (1) After the MRI is activated,  MHD turbulence develops near the equator.
 The magnetic field adds more pressure in the system which is  enhanced by 
 compression behind the shock and thus breaks the original equilibrium of the standing shock of the
 HD configuration.
 As a result, the centrifugally supported shock moves back and forth between
  60 $R_{\rm g} \leq R \leq 170 R_{\rm g}$.
  In addition to the outer shock, 
 another inner weak shock appears irregularly with rapid variations due to the
 interaction of the expanding high magnetic blob with the accreting matter below the outer shock.

 (2) This process repeats irregularly with an approximate  time-scale of  
  (4 -- 5) $\times 10^5$ s ($\sim$ 5 days) 
    with an accompanying smaller amplitude modulation with a period of $\sim 0.9\times 10^5$ s (25 hrs).
  The time-variability with two different periods  is attributed to the oscillating outer
  strong shock together with the rapidly oscillating  inner weak shock.
 Due to the variable shock location, the luminosities vary by more than a factor of $3$ and the
 average values are  3 -- 4 $\times 10^{34}$ erg s$^{-1}$.

 (3) The variability patterns of the order of $\sim$ 5 days and 25 hrs found in models A and B
   are compatible with the latest results of long-term flares of Sgr A* with X-ray luminosity
  $L_{\rm x} > 10^{34}$ erg s$^{-1}$ occurring  at a rate of $\sim1$ per day 
 and with luminous flares with $L_{\rm x} > 10^{35}$ erg s$^{-1}$ occurring approximately 
 every  5 -- 10 days by Chandra, Swift and XMM-Newton monitoring of Sgr A*.

 The time-scale of the irregular variability in the present model depends on the location of the centrifugally
 supported shocks in the accretion flow.
 The irregular variabilities in these models are due to the competition among the gravity, centrifugal force
 and pressure gradient forces at the shock, that is, the unstable behavior of  the standing shock. 
 It is known that a standing shock in adiabatic accretion flow is unstable against axisymmetric perturbations
 under some conditions and that the dynamical time-scale of the instability is of the order of
 $\sim R_{\rm s}/v_{\rm -}$, where $v_{\rm -}$ is the pre-shock velocity \citep{key-37}.
Therefore, the smaller the shock position $R_{\rm s}$  the faster the shock variability is.
 If we had started with initial conditions in the magnetized flow leading to a smaller shock location,
 a more rapid variability could be obtained, and vice versa.
 Accordingly, the inclusion of other physical processes like viscosity \citep{key-14},  the magnetic resistivity, 
 or other radiative cooling mechanisms \citep{key-9}, like synchrotron radiation in a stronger magnetic field 
 and different configurations of the magnetic field
  may change the present findings.

 The observed  rapid flares of Sgr A* with time-scales of hours to day show time lags of flares between 
 multi-wavelength bands, such as the 20 -- 40 minutes lag between 22 GHz and 43 GHz and about
 two hours lag between X-ray and submilimeter in Sgr A* (Yusef-Zadeh et al. 2008, 2009). 
 At present, the time lags of the long-term flares, for example,  between the  X-ray and radio 
 wavelengths have not been observed yet because simultaneous long time series at such wavelengths
 are limited.  However, it is natural for us to consider there exist the time lag phenomena of the long-term 
 flares. Such time lags may be caused by different physical processes or in different regions
 of the accretion inflow-outflow system.
 In this connection, we may need  an additional picture to describe the flare events of Sgr A*. 
 Here, we regard the average emission obtained in the models as the nearly constant emission of 
 the quiescent state. On the other hand, the variable emission in Sgr A* may be probably caused by 
 the intermittent outflows, as found by the present results.
 We can speculate that magnetized clumps from the shocked material near $z \sim 200 R_{\rm g}$ 
 are swept by the high speed jet.
  We then argue that the impact between the clumps could produce  a radio flare with a time lag 
 of $200 R_{\rm g}/c \sim $ 2 hrs.
 The radio emission could be  observed as synchrotron radiation in the surrounding of the clumps 
 if  electrons are relativistically accelerated by the enhanced magnetic field via Fermi shock or magnetic
 reconnection acceleration \citep{key-16,key-17,key-28,key-56}.

  The present low angular momentum flow model is too simple to reproduce the detailed observed
 spectra of Sgr A*.  With the inclusion of the physical processes described above and a longer term MHD simulation
in a larger computational domain will allow us to track the full evolution of this system and the different regimes 
 of the variability phenomena in low angular momentum flows.

 We would like to thank an anonymous referee for many useful comments.
C.B.S. acknowledges support from the Brazilian agency FAPESP (grant 2013/09065-8)
 and was also supported by the I-Core center of excellence of the CHE-ISF.
This work has made use of the computing facilities of the Laboratory of Astroinformatics (IAG/USP, NAT/Unicsul), 
whose purchase was made possible by FAPESP (grant 2009/54006-4).
AN thanks GD, SAG; DD, PDMSA and Director, URSC for encouragement and
continuous support to carry out this research.  
EMDGP  acknowledges partial support from the Brazilian agencies FAPESP (2013/10559-5) and CNPq (306598/2009-4) grants.

\end{document}